\documentclass[prl,twocolumn,aps,superscriptaddress,notitlepage,floatfix,10pt]{revtex4-2}
\usepackage{amsmath,mathtools,amsthm,amssymb,pifont}
\usepackage[percent]{overpic}
\usepackage[utf8]{inputenc}
\usepackage{comment}
\usepackage[american]{babel}
\usepackage{graphicx,xcolor,bbold,titlesec}

\usepackage{MnSymbol}

\usepackage[colorlinks,
bookmarksopen,
bookmarksnumbered,
citecolor=teal,
linkcolor=teal,
urlcolor=teal]{hyperref}
\usepackage{braket}
\usepackage{enumitem}
\usepackage{subfigure}
\usepackage{ifthen}
\usepackage{dsfont}

\usepackage{orcidlink}

\usepackage{rotating}
\usepackage{multirow}

\definecolor{lblue} {RGB}{51,71,158}

\begin{document}

\title{Fermionic magic resources in disordered quantum spin chains}

\author{Pedro R. Nic\'acio Falc\~ao\orcidlink{0000-0001-5893-0460}}
\email{pedroruyter.falcao@gmail.com}
\affiliation{Szkoła Doktorska Nauk \'Scis\l{}ych i Przyrodniczych, Uniwersytet Jagiello\'nski,  \L{}ojasiewicza 11, PL-30-348 Krak\'ow, Poland}
\affiliation{Instytut Fizyki Teoretycznej, Wydzia\l{} Fizyki, Astronomii i Informatyki Stosowanej,
Uniwersytet Jagiello\'nski,  \L{}ojasiewicza 11, PL-30-348 Krak\'ow, Poland}
\author{Jakub Zakrzewski\orcidlink{0000-0003-0998-9460}}
\email{jakub.zakrzewski@uj.edu.pl}
\affiliation{Instytut Fizyki Teoretycznej, Wydzia\l{} Fizyki, Astronomii i Informatyki Stosowanej,
Uniwersytet Jagiello\'nski,  \L{}ojasiewicza 11, PL-30-348 Krak\'ow, Poland}
\affiliation{Mark Kac Complex Systems Research Center, Uniwersytet Jagiello{\'n}ski, Krak{\'o}w, Poland}
\author{Piotr Sierant\orcidlink{0000-0001-9219-7274}}
\email{piotr.sierant@bsc.es}
\affiliation{Barcelona Supercomputing Center, Barcelona 08034, Spain}

\date{\today}

\begin{abstract}
Fermionic non-Gaussianity quantifies a quantum state’s deviation from a classically tractable free-fermionic description, constituting a necessary resource for computational quantum advantage. Here we use fermionic antiflatness (FAF) to measure this deviation across ergodic and many-body localized (MBL) regimes.
We focus on the paradigmatic disordered spin-$1\!/2$ XXZ chain and its impurity variant with local interactions. Across highly excited eigenstates, FAF evolves from typical-state behavior at weak disorder to strongly suppressed values deep in the MBL regime, with volume-law scaling in the XXZ chain and an area-law bound in the impurity setting. 
Rare long-range cat-like eigenstates exhibit a pronounced enhancement of FAF, making it a sensitive diagnostic of mechanisms proposed to destabilize MBL.
Starting from product states, we find that in the MBL regime FAF grows slowly in time, approaching saturation via a power-law relaxation. 
Overall, our results show that MBL suppresses fermionic non-Gaussianity, and the associated complexity beyond free fermions, while ergodicity restores it, motivating explorations of fermionic non-Gaussianity in other ergodicity-breaking phenomena.
\end{abstract}

\maketitle

\paragraph{Introduction.}  
Quantum resource theories~\cite{Chitambar19} provide a framework to characterize nonclassical features of quantum states by specifying a set of free states, operations,  and monotones that quantify deviations from them. 
When the free set is efficiently classically simulable, resource measures quantify classical description complexity and diagnose quantum computational advantage~\cite{Preskill12frontier, Daley22practical}.
A canonical example is the entanglement theory~\cite{Horodecki09quantum}, where separable states are free and weakly entangled states admit efficient tensor-network representations~\cite{Vidal03a, Verstraete04, Verstraete08, Orus14, Haegeman16, Orus19, ran2020tensor, Banuls23}. 
Yet, entanglement alone does not guarantee quantum advantage: even highly entangled stabilizer states~\cite{Smith06typical, Dahlsten07distr} remain efficiently simulable~\cite{Gottesman98,Aaronson04improved}, prompting the resource theory of non-stabilizerness (“magic”)~\cite{Veitch12, Veitch14stab, Howard17rom, Wang19chan, Liu22many, Leone22sre}.

Fermionic Gaussian states form another physically important class of efficiently simulable many-body states~\cite{Valiant01,Terhal02,Bravyi05flo}.
They arise from evolutions generated by Hamiltonians quadratic in Majorana operators~\cite{Majorana2006,Bravyi05flo}, or equivalently from matchgate circuits~\cite{Valiant01,Terhal02,Jozsa08}.
The covariance-matrix formalism enables efficient classical simulation of fermionic Gaussian states and provides access to equilibrium and nonequilibrium physics in a broad class of free-fermion models, including the transverse-field Ising and XY chains~\cite{Surace22, Mbeng24}.
Interactions and other non-Gaussian operations drive the state away from the Gaussian manifold~\cite{Paviglianiti25emergence}, hindering efficient classical simulation~\cite{Oszmaniec22, Boutin21, ReardonSmith24improved, Dias24classical}. This motivates the resource theory of fermionic non-Gaussianity, also termed fermionic magic resources~\cite{Hebenstreit19, Cudby24gaussian,Reardon24extent,Bittel24optimal,Hakkaku22}, with fermionic Gaussian states as the free set.
Measures of fermionic non-Gaussianity~\cite{Hebenstreit19,Cudby24gaussian, Bittel24optimal, Reardon24extent, Gottlieb05, Gottlieb07, Lumia24, Lyu24NGE, Coffman25magic, Bellomia25correlation} include fermionic antiflatness (FAF)~\cite{Sierant26faf}, argued recently to be both accessible in numerics and amenable to experimental measurements.

\begin{figure}[t]
    \centering
    \includegraphics[width=1\linewidth]{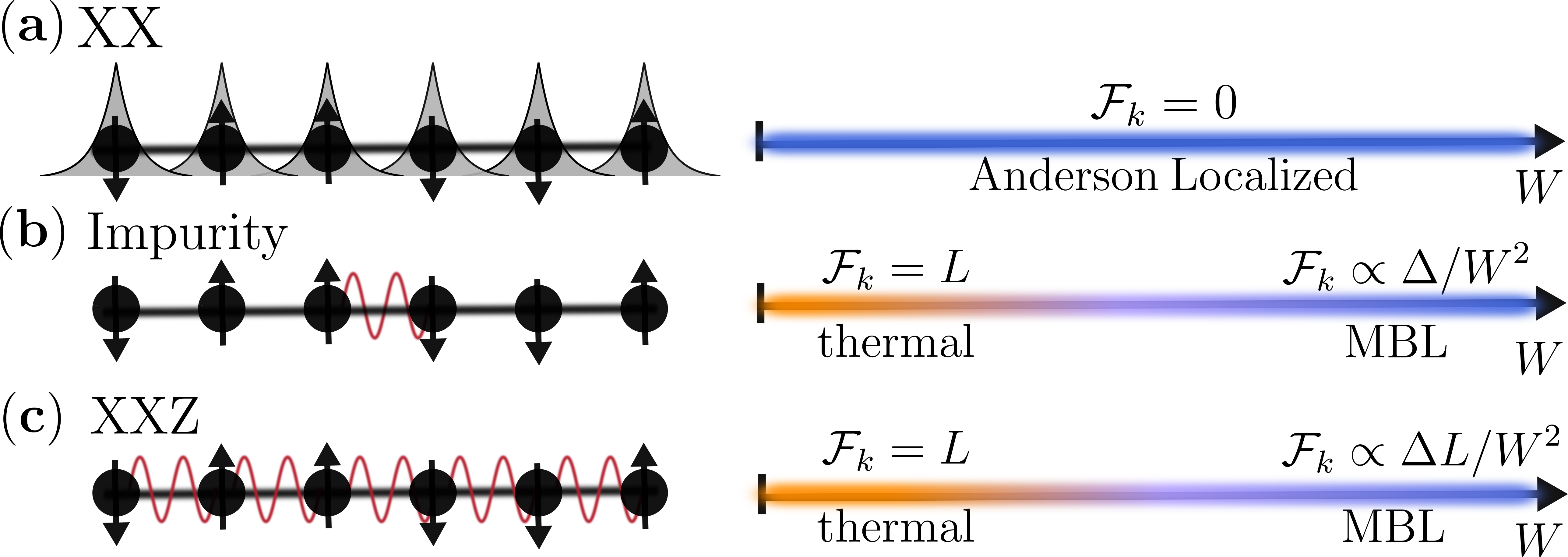}
    \caption{(a) The disordered XX chain is Anderson localized and fermionic Gaussian, hence \(\mathcal{F}_k=0\). Adding interactions \(\Delta\ne 0\) gives (b) an impurity model and (c) the XXZ chain, featuring ergodic and MBL regimes with non-Gaussian eigenstates and \(\mathcal{F}_k>0\), decaying with disorder strength $W$ as \(\mathcal{F}_k\propto W^{-2}\).}
    \label{Cartoon}
\end{figure}

Entanglement has long served as a central diagnostic of quantum many-body ground states~\cite{Amico08entanglement, Eisert10area, Laflorencie16, Calabrese04} and nonequilibrium dynamics~\cite{Eisert15,Calabrese09entanglement,Abanin19}. More recently, attention has increasingly turned to non-stabilizerness, with intensive studies across equilibrium phases and quantum phase transitions~\cite{White21, haug2023quantifying, tarabunga23gauge, Odavic23,  Passarelli24, Tarabunga24rk, Falcao25, Jasser2025, Bera2025SYK} and in out-of-equilibrium settings~\cite{rattacaso2023stabilizer, Turkeshi25spectrum, turkeshi2024magic, Tirrito24anti, Odavic2025, Haug2025probingquantum, kos2024exact, Santra25complexity, bejan2025magic, tirrito2025universal, Passarelli25, Hou2025highway, falcao2025nonstabilizerness,Hernandez26, Loio26telep}.
By comparison, studies of fermionic non-Gaussianity in many-body systems remain relatively scarce~\cite{Sierant26faf,Santra25abelian}.

Ergodic many-body systems thermalize~\cite{Rigol_2008,Dalessio16}, dispersing local information from the initial state into entanglement and correlations throughout the system. 
Thermalization is accompanied by a rapid generation of fermionic non-Gaussianity~\cite{Sierant26faf, Aditya25growth, Aditya2025mpemba}. 
In this Letter, we use fermionic antiflatness (FAF) to investigate how many-body localization (MBL)--a paradigmatic and robust mechanism of ergodicity breaking~\cite{Nandkishore15,Alet18,Sierant25rev}--hinders the buildup of fermionic magic resources. 
We study the disordered spin-$1\!/2$ XXZ chain and an impurity model with local interactions (Fig.~\ref{Cartoon}), which provide complementary routes from an Anderson-localized free-fermion limit to ergodic and MBL regimes. 
Investigating highly excited eigenstates, which underpin eigenstate-based characterizations of nonequilibrium behavior, and time evolution from product initial states, we show how disorder and interactions shape fermionic non-Gaussianity across the ergodic and MBL regimes.

\paragraph{Free fermions and fermionic non-Gaussianity.}
We study spin-$1\!/2$ chains of the form $\hat{\mathcal H}=\hat{\mathcal H}_{\rm xx}+\hat{\mathcal H}_{\mathrm{int}}$, where
\begin{equation}
\hat{\mathcal H}_{\rm xx}
=\frac{1}{4}\sum_{j=1}^{L}\!\Big(\hat{\sigma}^x_j \hat{\sigma}^x_{j+1}+\hat{\sigma}^y_j \hat{\sigma}^y_{j+1}\Big)
+\frac{1}{2}\sum_{j=1}^{L} h_j \hat{\sigma}^{z}_j \,,
\label{Hamilt_XX}
\end{equation}
$L$ is the system size, $\hat{\sigma}^{\alpha}_j$ (with $\alpha \in \{x,y,z\}$) are Pauli operators, $\hat{\mathcal H}_{\mathrm{int}}$ is the interaction term, and the onsite fields $h_j$ are drawn independently from a uniform distribution on $[-W,W]$, with $W$ the disorder strength.
We consider the disordered XXZ chain, a paradigmatic model for MBL studies~\cite{LFSantos04, Oganesyan07, Pal10, DeLuca13, Luitz15, Panda19, Suntajs20}, with Hamiltonian $\hat{\mathcal H}_{\rm xxz}=\hat{\mathcal H}_{\rm xx}+\hat{\mathcal H}_{\rm ext}$ and interaction term $\hat{\mathcal H}_{\rm ext}=\frac{\Delta}{4}\sum_{j=1}^{L}\hat{\sigma}^{z}_j\hat{\sigma}^{z}_{j+1}$, where $\Delta$ sets the interaction strength. 
In addition, we study an \textit{impurity model} $\hat{\mathcal H}_{\rm imp}=\hat{\mathcal H}_{\rm xx}+\hat{\mathcal H}_{\rm loc}$ with a single interacting bond, $\hat{\mathcal H}_{\rm loc}=\frac{\Delta}{4}\,\hat{\sigma}^{z}_{L/2}\hat{\sigma}^{z}_{L/2+1}$, confined to the center of the chain.

The Jordan--Wigner mapping~\cite{Jordan28} introduces a set of $2L$ Majorana operators
\begin{equation}
\hat\gamma_{2j-1}=\Big(\prod_{m<j} \hat{\sigma}^z_m\Big)\hat{\sigma}^x_j\,,
\qquad
\hat\gamma_{2j}=\Big(\prod_{m<j} \hat{\sigma}^z_m\Big)\hat{\sigma}^y_j\,,
\label{eq:WJ}
\end{equation}
which satisfy $\{\hat\gamma_m,\hat\gamma_n\}=2\delta_{mn}$. In terms of these operators, the free-fermion character of $\hat{\mathcal H}_{\rm xx}$ becomes explicit: it admits the quadratic Majorana form
$
\hat{\mathcal H}_{\rm xx}=\frac{i}{4}\sum_{m,n=1}^{2L}A^{\rm xx}_{mn}\,\hat\gamma_m\hat\gamma_n,
\label{eq:Hxx_quadratic}
$
where $A^{\rm xx}$ is a $2L \times 2L$ antisymmetric, $(A^{\rm xx})^{T} = - A^{\rm xx} $,  matrix~\footnote{Explicit expression for $\hat{\mathcal H}_{\rm xx}$ in terms of Majorana fermions is given in End Matter.}.
Any Hermitian operator \emph{quadratic} in Majoranas, $\hat K=\frac{i}{4}\sum_{m,n=1}^{2L}K_{mn}\,\hat\gamma_m\hat\gamma_n$ (with $K^{T}=-K$), generates a \emph{fermionic Gaussian unitary} $\hat U_K=\exp(-i\hat K)$~\cite{Bravyi05flo}. 
Under $U_K$, Majorana operators transform covariantly by an orthogonal rotation,
$\hat U_K^\dagger \hat \gamma_m \hat U_K=\sum_{n=1}^{2L}G_{mn}\hat\gamma_n,$ with $G=\exp(K)\in{\rm SO}(2L).$
Choosing $G$ that brings $A^{\rm xx}$ to its canonical block-diagonal form~\cite{Williamson36,Zumino62} yields the normal-mode decomposition
$\hat{\mathcal H}_{\rm xx}=\sum_{\alpha=1}^{L}\varepsilon_\alpha \hat Q_\alpha$ with $\hat Q_\alpha=\tfrac{i}{2}\tilde\gamma_{2\alpha-1}\tilde\gamma_{2\alpha}$ and $\tilde\gamma_m=\sum_n G_{mn}\hat\gamma_n$. 
Since the operators $\hat Q_\alpha$ mutually commute, the eigenstates of $\hat{\mathcal H}_{\rm xx}$ are simultaneous eigenstates of $\{\hat Q_\alpha\}$. Equivalently, they can be written as $\hat U_K^\dagger\ket{\mathbf z}$, where $\ket{\mathbf z}\in\mathcal B$ is a $\hat{\sigma}^z$-basis product state.
Likewise, $\ket{\psi(t)}=e^{-i\hat{\mathcal H}_{\rm xx}t}\ket{\mathbf z}$ remains fermionic Gaussian since $e^{-i\hat{\mathcal H}_{\rm xx}t}$ is itself a fermionic Gaussian unitary.

Fermionic Gaussian states obey Wick's theorem~\cite{negele2018quantum}: all higher-order correlators factorize into sums of products of two-point functions. 
Accordingly, a fermionic Gaussian state $\ket{\psi}$ is fully specified by its covariance matrix
$M_{mn}\equiv -\tfrac{i}{2}\bra{\psi}[\hat\gamma_m,\hat\gamma_n]\ket{\psi}$,
a real antisymmetric $2L\times2L$ matrix, so storing $\ket{\psi}$ reduces to storing $M$. 
Moreover, Gaussian unitaries act as orthogonal rotations of the Majorana operators, implying that $\ket{\psi} \mapsto \hat U_G \ket{\psi}$ corresponds to the covariance-matrix update
$M\mapsto \tilde M = G M G^{T}$.
Hence, simulating free-fermionic states reduces to operations on $M$, with classical resources scaling only polynomially with system size $L$.

The interaction terms $\hat{\mathcal H}_{\rm ext}$ and $\hat{\mathcal H}_{\rm loc}$ are \emph{quartic} in Majorana operators and thus break the free-fermion solvability of $\hat{\mathcal H}_{\rm xx}$, driving states away from the fermionic Gaussian manifold. 
To quantify the fermionic non-Gaussianity of state $\ket{\Psi}$, we use the FAF~\cite{Sierant26faf},
\begin{equation}
    \mathcal{F}_{1}(\ket{\Psi}) = L - \tfrac{1}{2}\mathrm{tr}\big[(M^TM)\big]\,, 
    \label{FAF}
\end{equation}
where $M$ is the covariance matrix~\footnote{In general, $\mathcal{F}_k(\ket{\Psi}) = L - \tfrac{1}{2}\mathrm{tr}\big[(M^TM)^k\big]$, for integer $k\geq1$. In this letter, we restrict ourselves to $k=1$ for concreteness.}. 
The FAF has the key properties expected of a non-Gaussianity measure~\cite{Sierant26faf}: it is \emph{faithful}, $\mathcal F_{1}(\ket{\Psi})=0$ if and only if $\ket{\Psi}$ is fermionic Gaussian; it is \emph{Gaussian invariant}, $\mathcal F_1(\hat U_G\ket{\Psi})=\mathcal F_1(\ket{\Psi})$ for any fermionic Gaussian unitary $U_G$; and it is \emph{additive}, $\mathcal F_1(\ket{\Psi} \otimes \ket{\Phi}) = \mathcal F_1(\ket{\Psi})+ \mathcal{F}_1(\ket{\Phi})$ if $\ket{\Psi}$ or $\ket{\Phi}$ has a fixed fermionic parity $\mathcal{P}=\prod_j \hat{\sigma}^{z}_j$, and subadditive otherwise.
Since $\mathcal{F}_1$ is a simple function of the two-point Majorana correlators encoded in $M$, FAF can be efficiently calculated numerically and is naturally suited for experiments~\cite{Zhao21, Wan23, Denzler24, Majsak2025, Haug26practical}.

\begin{figure}[t!]
    \centering
    \includegraphics[width=1\linewidth]{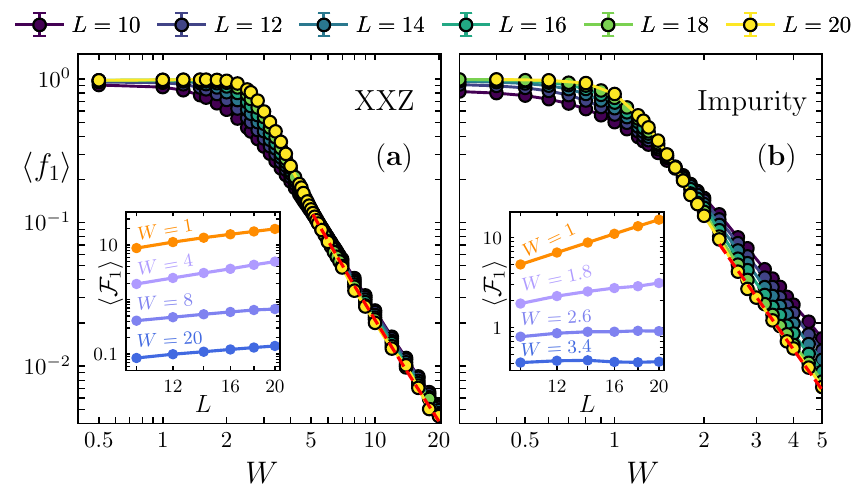}
    \caption{The FAF density $\langle f_1\rangle$ across the ergodic--MBL crossover in (a) the XXZ chain and (b) the impurity model ($\Delta=1$). At weak disorder $W$, $\langle f_1\rangle$ approaches its typical-state value; in MBL regime it decreases as $\langle f_1\rangle \propto (N_{\mathrm{int}}/L) W^{-2}$, with a subleading $W^{-3}$ correction (red dashed lines). Insets: $\langle \mathcal{F}_1\rangle$ as a function of $L$ at fixed $W$, showing that the volume law persists up to $W=20$ in the XXZ chain (a), while the impurity model (b) becomes $ L$-independent at larger $W$.}
    \label{FAF_density}
\end{figure}

\paragraph{Ergodic-MBL crossover.}
 
We employ the POLFED algorithm~\cite{Sierant20p} to compute $\min\{100,\mathcal N/20\}$ mid-spectrum eigenstates of $\hat{\mathcal H}_{\rm xxz}$ and $\hat{\mathcal H}_{\rm imp}$ in the zero-magnetization sector of dimension $\mathcal N=\binom{L}{L/2}$. 
{We average $\mathcal{F}_1$ over these eigenstates and over $10^3$--$10^5$ disorder realizations $\{h_j\}$~\cite{supmat}}. The resulting FAF density $\langle f_1\rangle=\langle\mathcal F_1\rangle/L$ is shown in Fig.~\ref{FAF_density}.

In the ergodic regime (weak disorder $W$), $\langle f_1\rangle$ tends to unity with increasing $L$, consistent with the typical-state value $\mathcal F_1^{\rm {Haar} } \simeq L$~\cite{Sierant26faf}.
In contrast, for large $W$, when the systems are in the MBL regime,
the FAF in eigenstates is limited, similarly to entanglement~\cite{Bauer13area, Serbyn13b, Huse14} and non-stabilizerness measures~\cite{falcao2025nonstabilizerness}.
{The FAF density as a function of $W$ exhibits a crossing point that separates the ergodic and MBL regimes. For the XXZ model, the crossing point is remarkably stable for different system sizes (see End Matter) and occurs at $W\approx 5$ for $L\geq15$. In contrast, standard ergodicity breaking indicators, such as the average gap-ratio~\cite{Oganesyan07}, exhibit strong finite-size drifts, and only their extrapolation yields $W\sim5$~\cite{Sierant20p, niedda25r}.
For the impurity model, the crossing point occurs at $W\approx1.6$~\footnote{Detailed analysis of ergodic-MBL crossover in the impurity model with standard ergodicity breaking indicators is performed in~\cite{PedroToAppear}}}.

The behavior of $\langle f_1\rangle$ in the MBL regimes of $\hat{\mathcal H}_{\rm xxz}$ and $\hat{\mathcal H}_{\rm imp}$ is markedly different. In the XXZ chain, $\mathcal{F}_1 \propto L$ yielding practically overlapping $\langle f_1 \rangle$ curves. By contrast, in the impurity model $\mathcal F_1$ becomes independent of $L$ at large $W$ [inset of Fig.~\ref{FAF_density}(b)].   This suggests that, in the MBL regime, the scaling of $\mathcal F_1$ is controlled by the spatial extent of interactions: extensive $\hat{\mathcal H}_{\rm ext}$ yields extensive FAF, whereas local $\hat{\mathcal H}_{\rm loc}$ yields a constant bound on $\mathcal F_{1}$.

\paragraph{Fermionic non-Gaussianity in the MBL regime. }

The disordered XX chain $\hat{\mathcal H}_{\rm xx}$ is a 1D free-fermion system and is Anderson localized~\cite{Anderson58,Evers08}. 
Accordingly, its normal-mode operators $\hat Q_\alpha$ are localized with localization length $\xi$ and have dominant support on single-site operators $\hat{\sigma}^{z}_j$, up to corrections decaying exponentially with distance. Since $[\hat{\mathcal H}_{\rm xx},\hat Q_\alpha]=0$, we view $\hat{Q}_\alpha$ as Anderson local integrals of motion (ALIOMs). 
Following~\cite{Vidmar21,Krajewski22Restoring}, we quantify the component of the interaction that genuinely breaks the free-fermionic structure by decomposing $\hat{\mathcal H}_{\rm int}$ into parts parallel and orthogonal to the ALIOM space with respect to the Hilbert--Schmidt product $\langle \hat A,\hat B\rangle \equiv \frac{1}{\mathcal N}\mathrm{tr}[\hat A^\dagger \hat B]$ (and norm $\|\hat A\|=\langle \hat A,\hat A\rangle^{1/2}$): $\hat{\mathcal H}_{\rm int}=\hat{\mathcal H}_{\rm int}^{\parallel}+\hat{\mathcal H}_{\rm int}^{\perp}$, with $\langle \hat{\mathcal H}_{\rm int}^{\parallel},\hat{\mathcal H}_{\rm int}^{\perp}\rangle=0$ and $\langle \hat{\mathcal H}_{\rm int}^{\perp},\hat Q_\alpha\rangle=0$.
The parallel component $\hat{\mathcal H}_{\rm int}^{\parallel}=\sum_\alpha \langle \hat{\mathcal H}_{\rm int},\hat Q_\alpha\rangle\,\hat Q_\alpha$ produces only diagonal energy shifts, whereas $\hat{\mathcal H}_{\rm int}^{\perp}$ acts as a nontrivial perturbation that breaks the fermionic Gaussianity of Anderson insulator. 
In the large-disorder limit, the leading contribution to $\|\hat{\mathcal H}_{\rm int}^{\perp}\|$ scales as $\|\hat{\mathcal H}_{\rm int}^{\perp}\|\sim \Delta N_{\rm int}/W$, where $N_{\rm int}=1$ for the impurity model and $N_{\rm int}=L$ for the XXZ chain, while $\|\hat{\mathcal H}_{\rm xx}\|\sim W$; {see~\cite{supmat} for details.}

Deep in the MBL regime, the Anderson-localized structure of $\hat{\mathcal H}_{\rm xx}$ persists in the interacting system in the form of quasi-local integrals of motion (LIOMs)~\cite{Huse14,Serbyn13b,Ros15}. Equivalently, there exists a quasi-local unitary $\hat U_M$ that adiabatically dresses the Anderson eigenstates $\ket{\Psi^{\rm And}}$ into the interacting ones $\ket{\Psi^{\rm MBL}}=\hat U_{M}\ket{\Psi^{\rm And}}$. 
Writing $\hat U_{M}=e^{\hat S}$ with $\hat S^\dagger=-\hat S$ quasi-local, {we split the generator as $\hat S=\hat S_2+\hat S_4$ according to the number of quasiparticle excitations it creates above $\ket{\Psi^{\rm And}}$, and denote by $\hat\Pi_k$ the projector onto the $k$-excitation sector. Only $\hat S_4$ leaves the Gaussian manifold, and its typical magnitude, set by the four-quasiparticle matrix elements of $\hat{\mathcal{H}}^{\perp}_{\rm int}$ and the corresponding energy denominators, scales as $\|\hat S_4\|\propto\Delta N_{\rm int}/W^{3}$ (see Ref.~\cite{supmat}).} The covariance matrix in $\ket{\Psi^{\rm MBL}}$ state can be obtained from the Anderson limit $M^{\rm And}_{mn}=-\tfrac{i}{2}\bra{\Psi^{\rm And}}[\hat\gamma_m,\hat\gamma_n]\ket{\Psi^{\rm And}}$, by conjugation of Majorana operators $\hat\gamma_m  \mapsto \hat U_{M}^{\dagger} \hat\gamma_m \hat U_{M}$. For $W\gg 1$, quasi-locality of $\hat{U}_M$ implies a controlled expansion $\hat\gamma_m \mapsto \hat\gamma_m+[\hat\gamma_m,\hat S]+O(\hat S^2)$,
yielding $M=M^{\rm And}+\delta M$, where $\delta M$ encodes the dressing induced by $\hat{\mathcal H}_{\rm int}^\perp$. Since the Anderson eigenstate is fermionic Gaussian, one has $(M^{\mathrm{And}})^{T}M^{\mathrm{And}} = \mathds{1}$~\cite{Bravyi05flo}, yielding, deep in the MBL regime,
\begin{equation}
    \mathcal{F}^{\rm typ}_1(\ket{\Psi^{\rm MBL}}) \simeq 16\lVert\hat{\Pi}_{4}\hat{S}_{4}\ket{\Psi^{\rm And}}\rVert^{2}\,,
    \label{Eq:F1_Scaling}
\end{equation}
{where the two-quasiparticle sector cancels at $O(\lVert\hat S\rVert^2)$~\cite{supmat}, leaving only the four-quasiparticle component of the dressing and yielding $\mathcal{F}_1^{\rm typ} \propto \Delta^2N_{\rm int}^{2}/W^{6}$.}

{Equation~\eqref{Eq:F1_Scaling} describes \emph{typical} MBL eigenstates, i.e., those for which the quasi-local expansion is controlled, and the median indeed follows $\mathcal{F}_1^{\rm typ}\propto \Delta^2 N_{\rm int}^{2}/W^6$ (see End Matter). The mean plotted in Fig.~\ref{FAF_density}, on the other hand, decays only as $\langle \mathcal{F}_1\rangle\propto  N_{\rm int}/W^2$. The mean is therefore not set by the typical dressing, but is dominated by rare eigenstates in which the controlled expansion breaks down.} 

\begin{figure}[t!]
    \centering
    \includegraphics[width=1\linewidth]{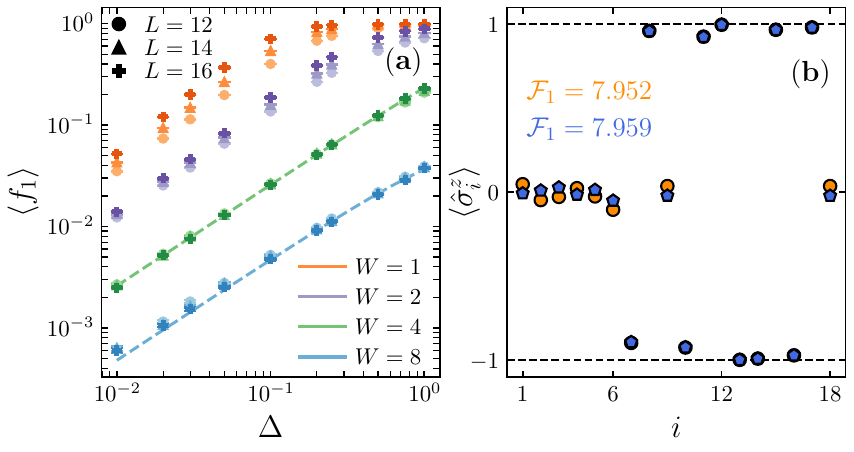}
    \caption{(a) The FAF density $\langle f_1\rangle$ as a function of interaction strength $\Delta$ in XXZ model for disorder strength $W$ and system size $L$, the dashed lines show fits $\braket{f_1}=a\Delta + b\Delta^2$, with $b\Delta^2$ being a small, sub-leading correction; (b) Expectation values $\langle \hat{\sigma}^{z}_i\rangle$ in two long-range cat-like eigenstates of $\hat{\mathcal H}_{\rm xxz}$ for $L=18$ and $W=14$, and the corresponding  $\mathcal{F}_1$ values.}    
    \label{Fig:LR_Cat}
\end{figure}

Figure~\ref{Fig:LR_Cat}(a) shows $\langle f_1\rangle$ versus interaction strength $\Delta$ for several disorder values $W$. At large $W$, $\langle f_1\rangle \propto \Delta$ with high accuracy. {This linear dependence corroborates the same conclusion: a smooth dressing would inherit the $\Delta^2$ scaling, whereas a resonant eigenstate carries a FAF bounded by the number of participating spins, decaying as a Lorentzian in the detuning over a window $\Delta$~\cite{supmat}. Averaging this bounded observable yields a mean linear in $\Delta$; its $W^{-2}$ decay is set by the resonance probability~\cite{supmat}.}
In contrast, at weak disorder (e.g., $W=1$) even at small fixed $\Delta$, we observe a pronounced growth of $\langle f_1\rangle$ with system size $L$, signaling an instability of the Anderson insulator and the onset of ergodicity. In the thermodynamic limit, this behavior is consistent with a nonanalytic dependence on $\Delta$, with FAF jumping from $\mathcal F_1=0$ at $\Delta=0$ to an extensively scaling value set by the typical-state behavior $\langle f_1\rangle\simeq 1$.

\paragraph{Cat-like eigenstates.}
While mathematical constructions~\cite{Imbrie16a,deRoeck24absence} support the existence of a quasilocal dressing unitary $\hat U_M$ at sufficiently large disorder $W$ (even in the thermodynamic limit), the fate of the ergodic--MBL crossover as $L\to\infty$ remains debated~\cite{Sierant25rev}. 
A widely discussed mechanism for destabilizing MBL involves rare many-body resonances~\cite{Villalonga20,Garratt21Local,Crowley21,colbois24interaction,Colbois24a,laflorencie25cat,morningstar22landmarks,Ha23, Biroli24, Miranda26}, which evade a simple adiabatic connection to Anderson eigenstates. 
In this scenario, interactions hybridize two spin configurations that differ over a macroscopic region, producing a cat-like superposition on that segment while the remaining degrees of freedom stay effectively frozen~\cite{laflorencie25cat}. 
Such resonant eigenstates can occur even deep in the MBL regime and are visible in spin correlations~\cite{laflorencie25cat,Padhan2026}.

In Fig.~\ref{Fig:LR_Cat}(b) we show the profiles $\langle \hat{\sigma}^{z}_i\rangle$ for a pair of representative long-range resonant eigenstates deep in the MBL regime. 
The $\braket{\hat{\sigma}^{z}_i}$ profiles suggest an ansatz~\cite{laflorencie25cat}
\begin{equation}
|\Psi^{\pm}_{\rm cat}\rangle=|\varphi\rangle\otimes|\Psi_{\rm fr}\rangle\,,
\label{MBL_Cats}
\end{equation}
where $\ket{\Psi_{\rm fr}}$ is a product state on the nonresonant sites with $\langle \hat{\sigma}^{z}_i\rangle=\pm 1$, and
$\ket{\varphi}=\tfrac{1}{\sqrt{2}}\big(\ket{\uparrow,\downarrow,\uparrow,\downarrow,\ldots} \pm \ket{\downarrow,\uparrow,\downarrow,\uparrow,\ldots}\big)$
is a cat superposition on the resonant segment. 
Due to FAF additivity, $\mathcal F_1(\ket{\Psi^{\pm}_{\rm cat}})=\mathcal F_1(\ket{\varphi})+\mathcal F_1(\ket{\Psi_{\rm fr}})$. Since $\ket{\Psi_{\rm fr}}$ is fermionic Gaussian (hence $\mathcal F_1(\ket{\Psi_{\rm fr}})=0$),  FAF isolates the contribution of the cat component. 
The cat state $\ket{\varphi}$ has vanishing covariance matrix, and $\mathcal F_1(\ket{\varphi})$ is equal to the number of spins participating in the many-body resonance.
Consistently, for the states in Fig.~\ref{Fig:LR_Cat}(b) we find $\mathcal F_1\simeq 7.95$, in close agreement with the size of the resonant region inferred from $\langle \hat{\sigma}^{z}_i\rangle$.

These results show that many-body resonances, a leading candidate mechanism for destabilizing MBL, are accompanied by a pronounced enhancement of fermionic non-Gaussianity: cat-like resonant eigenstates exhibit much larger FAF than typical eigenstates at the same disorder, demonstrating the direct link between nonperturbative mechanisms destabilizing MBL and fermionic magic resources. 
At present, the impact of such rare resonances on the fate of MBL in the thermodynamic limit remains unresolved, since at numerically accessible system sizes long-range resonant states occur with low probability. 
In contrast, at the ergodic--MBL crossover (e.g., $W=5$ in the XXZ chain) the dominant resonances are typically short-ranged, involving $\sim4$ sites and producing a peak around $\mathcal F_{1}\simeq 4$ in the FAF distribution; a quantitative analysis of resonance statistics, including longer-range events {and how they dominate the mean FAF}, is presented in~\cite{supmat}.

\paragraph{Time dynamics.} Finally, we investigate the real-time dynamics of the FAF in the MBL regime. We compute the time-evolution operator $U(t)=e^{-i\hat{\mathcal H}t}$ by full exact diagonalization of $\hat{\mathcal H}_{\rm xxz}$ and $\hat{\mathcal H}_{\rm imp}$. We calculate the FAF of the time-evolved state $\ket{\Psi_t} = U(t) \ket{\Psi_0}$ with an initial N{\'e}el state $\ket{\Psi_0} = \ket{\uparrow\downarrow...\uparrow\downarrow}$, and average the results over $10^4$ disorder realizations to obtain the average  FAF, $\braket{\mathcal{F}_1(\ket{\Psi_t})}$. {Additional analysis for other disorder strengths and initial states can be found in~\cite{supmat}.}

\begin{figure}[t!]
    \centering
    \includegraphics[width=1\linewidth]{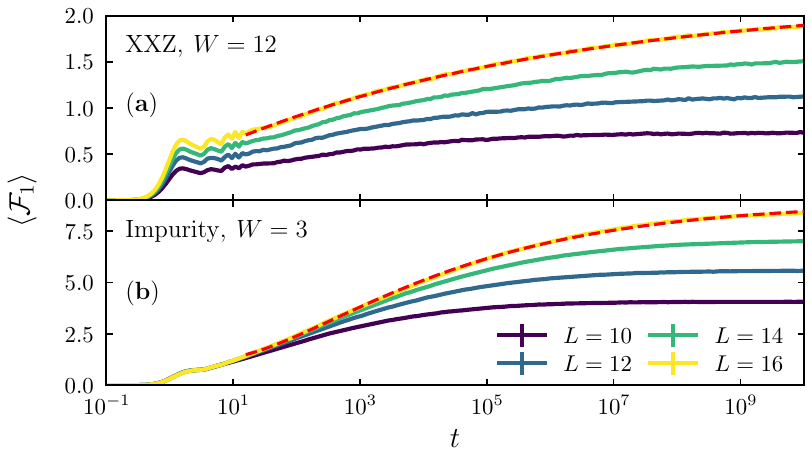}
    \caption{Time dynamics of $\langle \mathcal{F}_1\rangle$ for the (a) XXZ and (b) impurity model in the MBL regime, with interaction strength set as $\Delta=1$. Evolution of $\langle \mathcal{F}_1\rangle$  follows the prediction in Eq.~\eqref{TD_MBL}, as shown by the red-dashed lines; in the impurity model the second-order term $t^{-2\beta}$ is non-negligible.
    }
    \label{Fig:TD}
\end{figure}

The results for $\langle\mathcal F_1\rangle$ in the MBL regime of the XXZ chain are shown in Fig.~\ref{Fig:TD}(a) for system size $L=10,\ldots,16$. 
The initial N\'eel product state is fermionic Gaussian, hence $\langle\mathcal F_1(\ket{\Psi_0 })\rangle=0$. 
In contrast to ergodic dynamics, where FAF grows rapidly and saturates on timescales set by $L$~\cite{Sierant26faf}, in the MBL regime $\langle\mathcal F_1\rangle$ increases slowly over several decades in time, reminiscent of the slow entanglement growth phenomenology at strong disorder~\cite{DeChiara06,serbyn2013universal,Znidaric08,Znidarivc18e,Bardarson12}. 
An analogous slow growth is observed in the impurity model at $W=3$ [Fig.~\ref{Fig:TD}(b)]. 
At fixed intermediate times (e.g., $t=10$), the system-size dependence differs: in the XXZ chain $\langle\mathcal F_1\rangle$ grows approximately linearly with $L$, while in the impurity model it is nearly $L$ independent. 
Nevertheless, the long-time saturation value is extensive in $L$ in \emph{both} models, paralleling the volume-law saturation of entanglement entropy~\cite{DeChiara06,serbyn2013universal,Znidaric08,Znidarivc18e,Bardarson12}.

To understand the FAF dynamics in the MBL regime, we note that Eq.~\eqref{FAF} implies
\begin{equation}
\mathcal F_1(\ket{\Psi_t})= L-\sum_{m<n}\big|\langle \hat\gamma_m\hat\gamma_n\rangle_{t} \big|^2\,,
\label{eq:faf1}
\end{equation}
so $\langle\mathcal F_1\rangle$ is governed by two-point Majorana correlators $\langle \hat\gamma_m\hat\gamma_n\rangle_{t}$. 
Deep in MBL, these correlators exhibit slow relaxation. The onsite bilinears $\hat{\gamma}_m \hat{\gamma}_n$ with $(m,n)=(2k\!-\!1,2k)$ reduce{, up to a phase, to $i\hat \sigma^z_k$ on the same site $k$; hence $|\langle\hat{\gamma}_m\hat{\gamma}_n\rangle_t|^2 =|\langle\hat{\sigma}^{z}_{k}\rangle_t|^{2}$}, which approaches a constant $c_z\leq1$ with a power-law decay~\cite{Sierant25rev}. The nonlocal bilinears with $|m-n|>1$ map under Eq.~\eqref{eq:WJ} to $\hat \sigma^\alpha_i\big(\prod_{j=i+1}^{l-1}\hat \sigma^z_j\big)\hat  \sigma^{\alpha^{\prime}}_l$ {, where $i=\lceil m/2\rceil$ and $l=\lceil n/2\rceil$ label the two sites, and $\alpha,\alpha'\in\{x,y\}$ are fixed by the parities of $m$ and $n$. Their} expectation values approach a saturation value with a power-law correction $\sim t^{-A\beta}$, where {$A\in\{1,2\}$ distinguishes the regimes before and after sites $i$ and $l$ dephase, and $\beta\leq \xi\ln{2}$ is bounded by the $\ell$ bit localization length $\xi$~(see End Matter), and is correspondingly small.} 
{Splitting~\eqref{eq:faf1} by distance $|m-n|$: the $L$ onsite terms give $L-\sum_{k=1}^L |\braket{\hat\sigma_{k}^{z}}_t|^2$; the $O(L^2)$ nonlocal terms are individually small but positive, making the onsite piece a strict upper bound on $\mathcal{F}_1(\ket{\Psi_t})$, and their approach to saturation, described above, generates the power-law term in Eq.~\eqref{TD_MBL}. As a result,}
\begin{equation}
\mathcal F_1(\ket{\Psi_t})
=\mathcal F_1^{\rm MBL}-C\,t^{-\beta}+O(t^{-2\beta})\,,
\label{TD_MBL}
\end{equation}
{where $\mathcal F_1^{\rm MBL}\leq \sum_{k=1}^L (1-\braket{\hat\sigma_{k}^{z}}_{\infty}^2)$ is the saturation value and $C$ is a constant. The fits in Fig.~\ref{Fig:TD} for both the XXZ and impurity models are consistent with Eq.~\eqref{TD_MBL}. Additional discussion about the dynamics of the FAF is presented in~\cite{supmat}.}

\paragraph{Conclusion.} 

In this work we characterized fermionic magic resources in out-of-equilibrium disordered spin-$1\!/2$ chains. 
By analyzing highly excited eigenstates, we found that at strong disorder the MBL regime remains smoothly connected to the free-fermionic Anderson-insulator limit: fermionic non-Gaussianity is suppressed as disorder increases, and its scaling is set by the amplitude and spatial extent of the interactions.
The rare long-range cat-like resonance eigenstates, while having negligible weight in disorder-averaged observables at accessible sizes, exhibit a pronounced enhancement of fermionic non-Gaussianity, tying candidate nonperturbative mechanisms for destabilizing MBL to beyond free-fermion state complexity. 
At weak disorder, in contrast, interactions destabilize the Anderson limit and drive the system into an ergodic regime where fermionic non-Gaussianity approaches typical-state values with increasing system size. 
These eigenstate trends are mirrored in dynamics: deep in MBL, fermionic non-Gaussianity grows anomalously slowly and relaxes toward an extensive plateau via a power-law in time.

We conjecture that the phenomenology of fermionic magic resources established here extends to a broader class of systems that exhibit MBL and admit a well-defined localized free-fermion limit, including models with quasiperiodic potentials~\cite{Iyer13,Doggen19qp,Agrawal20,Aramthottil21,falcao2024many}, disorder-free localization~\cite{Schulz19,Nieuwenburg19,Taylor20,Chanda20,Yao20b,Yao21,Yao21b}, and disordered Ising chains~\cite{Kjall14}. 
In the End Matter we further support this conjecture by showing that the LIOM description~\cite{Serbyn13b,Huse14} reproduces the FAF behavior observed in our results. 
More broadly, our results provide a reference point for future studies of fermionic non-Gaussianity in other ergodicity-breaking phenomena, including disorder-induced regimes beyond simple LIOM phenomenology such as bond disorder~\cite{Aramthottil24} and random interactions~\cite{Li17}, 
quantum many-body scars~\cite{Serbyn21,Moudgalya22,Szoldra22,Aramthottil22,Chandran23quantum} and ergodicity breaking due to gauge constraints~\cite{Brenes18} or other conservation laws~\cite{Rakovszky20}. 
We leave for future work the question of how ergodicity breaking in these settings constrains fermionic magic resources and quantum-state complexity.

\paragraph{Acknowledgments.}
We acknowledge insightful discussions with Konrad Pawlik.
P.S. acknowledges collaboration with Sreemayee Aditya, Paolo Stornati, Emanuele Tirrito, and Xhek Turkeshi on related topics, and useful discussions with members of Quantic group at BSC. The work of P.R.N.F. and J.Z. was funded by the National Science Centre, Poland, project  2021/43/I/ST3/01142 -- OPUS call within the WEAVE programme. 
P.S. acknowledges fellowship within the “Generación D” initiative, Red.es, Ministerio para la Transformación Digital y de la Función Pública, for talent attraction (C005/24-ED CV1), funded by the European Union NextGenerationEU funds through PRTR. We gratefully acknowledge the Polish high-performance computing infrastructure PLGrid (HPC Centers: ACK Cyfronet AGH) for providing computer facilities and support within the computational grant no. PLG/2025/018400.

\let\origaddcontentsline\addcontentsline
\renewcommand{\addcontentsline}[3]{}
%

\let\addcontentsline\origaddcontentsline

\onecolumngrid 
\section*{End Matter} 
\twocolumngrid 

\paragraph{Crossing point analysis of the FAF.}~In the main text, we noted that the average FAF $\langle f_1\rangle$ crosses at a nearly stable disorder strength, with $W\approx5$ for the XXZ Hamiltonian. Standard ergodicity-breaking indicators, by contrast, show strong finite-size drifts as system size increases, challenging the stability of the MBL regime~\cite{Sierant25rev}. We now compare the scaling of the crossing points of $\langle f_1\rangle$ with those of the average gap ratio $\langle r\rangle$, highlighting that the finite-size drifts are much milder in the former quantity.

\begin{figure}
    \centering
    \includegraphics[width=1.0\linewidth]{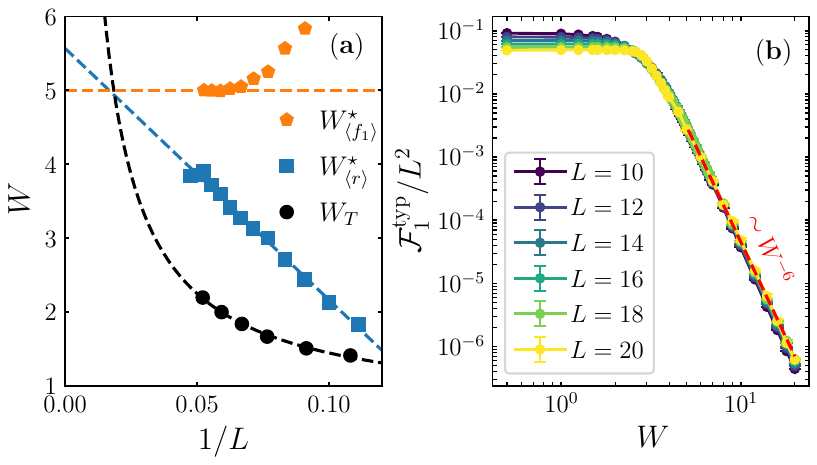}
    \caption{(a) Crossing point $W_{\langle f_1\rangle}^{\star}$ versus $1/L$ for the average FAF $\langle f_1\rangle$~(orange), compared to the gap-ratio crossing $W_{\langle r\rangle}^{\star}$~(blue) and deviations from the ergodic region $W_T$~(black); data taken from Ref.~\cite{Sierant20p}. The dashed lines are extrapolations to $L\rightarrow \infty$. $W_{\langle f_1\rangle}^{\star}\approx 5$ is nearly size-independent for $L\geq 15$, while $W_{\langle r\rangle}^{\star}$ and $W_T$ exhibits strong drifts. (b) Median FAF vs. $W$ for $L=10-20$: the red dashed line $\mathcal{F}_1^{\rm typ}\propto W^{-6}$, and the collapse under $\mathcal{F}_1^{\rm typ}/L^2$ confirms the $N_{\rm int}^2$ scaling of Eq.~(4).}
    \label{Fig:Crossing_TypScaling}
\end{figure}

We extract the disorder strength $W_{\langle f_1\rangle}^{\star}$ at which the curves of two different system sizes cross (see Ref.~\cite{supmat} for further details). The resulting $W_{\langle f_1\rangle}^{\star}$ [orange markers, Fig.~\ref{Fig:Crossing_TypScaling}(a)] stabilizes at $W\approx5$ for $L\geq15$, with markedly weaker drift than standard indicators. Interestingly, the crossing point approaches $W\approx5$ from above. On the other hand, we show in Fig.~\ref{Fig:Crossing_TypScaling}(a) the average gap-ratio crossing $W_{\langle r\rangle}^{\star}$~[blue squares] and its deviation from the GOE value $W_T$~[black circles]. The former locates the disorder strength at which the ETH-MBL crossover occurs, while the latter determines the value of $W$ at which the system no longer satisfies the ETH predictions. The disorder strengths exhibit strong finite-size drifts, with the extrapolation of their scalings meeting at $W\approx 5$.

The origin of the near-constancy of $W_{\langle f_1\rangle}^{\star}$ remains elusive. We stress, however, that this quantity is insensitive to single-particle effects and accounts only for genuine many-body correlations, which may be at the heart of such stability.

\paragraph{Typical scaling of the FAF in the MBL regime.}

In Eq.~\eqref{Eq:F1_Scaling} of the main text, we provide a closed-form expression for the typical FAF in the MBL regime~(Eq.~\eqref{Eq:F1_Scaling}; proof in Ref.~\cite{supmat}), yielding $\mathcal{F}_1^{\rm typ}\propto \Delta^2N_{\rm int}^2/W^6$ at $W\gg1$. The average value, on the other hand, is dominated by non-perturbative arguments and yields a different scaling, as we show in Fig.~\ref{FAF_density}. Here, we show numerically that the scaling of the typical behavior, $\mathcal{F}_1^{\rm typ}$, indeed follows the scaling predicted by Eq.~\eqref{Eq:F1_Scaling}.

Figure~\ref{Fig:Crossing_TypScaling}(b) shows the \emph{median} FAF over all mid-spectrum eigenstates: for $W\geq5$, it follows $\mathcal{F}_1^{\rm typ}\propto W^{-6}$, and the curves collapse under $\mathcal{F}_1^{\rm typ}/L^2$, confirming the $N_{\rm int}^2$ dependence. Additional numerical results are given in Ref.~\cite{supmat}.

\paragraph{Phenomenological description of MBL.}

Deep in the MBL regime, interacting disordered models can be described with good accuracy by an extensive set of quasi-local integrals of motion (LIOMs), also known as $\ell$-bits (localized bits)~\cite{Serbyn13b,Huse14,Ros15}. Within this phenomenological framework, the effective Hamiltonian takes the form
\begin{equation}
    \hat{\mathcal{H}}_{\mathrm{\ell-bit}}= \sum_{i} h_i\hat{\tau}_i^{z} + \sum_{i<j}J_{ij}\hat{\tau}_i^{z}\hat{\tau}_j^{z} + \sum_{i<j<k}J_{ijk}\hat{\tau}_i^{z}\hat{\tau}_j^{z}\hat{\tau}_k^{z} + ... \,,
    \label{eq:l-bit}
\end{equation}
where $h_i$ are on-site random energies, reminiscent of the normal-mode decomposition of the XX model [Eq.~\eqref{Hamilt_XX}], and the interactions decay exponentially with the distance between the $\ell$-bits, $J_{ij}\sim J_0e^{-|i-j|/\xi^{\prime}}$, with $\xi^{\prime}$ setting the interaction range and $J_0$ the interaction scale. The Hamiltonian~\eqref{eq:l-bit} captures several properties of the MBL regime~\cite{serbyn2014quantum, prakash21u, aceituno24a, Berger25, Szoldra24, Scocco24}, and constructive schemes to find the $\ell$-bits have been developed, e.g., in the XXZ chain~\cite{Chandran15,Kul18}. 

Here, we employ a $U(1)$-symmetric circuit $\hat{U}_\ell$ that maps the $\ell$-bit basis to the  
$\hat{\sigma}^z$-basis~\cite{aceituno24a}. The circuit is arranged in a brickwork structure, with a depth of $D_u$, where each two-site gate $\hat{U}_i$ is a random $4\times4$ Hermitian operator of the form $\hat{U}_i=e^{-ifw_i\hat{\Gamma}_i}$, with $w_i = e^{-2|h_i-h_{i+1}|}$ suppressing the spread of correlations, and
\begin{align}
    \hat{\Gamma}_i = \frac{\theta_1}{2}\hat{\sigma}^{z}_i + \frac{\theta_2}{2}\hat{\sigma}^{z}_{i+1} +\frac{\theta_3}{2}\lambda \hat{\sigma}^{z}_{i} \hat{\sigma}^{z}_{i+1}
    +c\hat\sigma_i^{+}\hat\sigma_{i+1}^{-} + c^{*}\hat\sigma_i^{-}\hat\sigma_{i+1}^{+}\,,
\end{align}
where $\hat{\sigma}_j^{\pm} = \tfrac{1}{2}(\hat{\sigma}_j^{x} \pm i\hat{\sigma}_j^{y})$, and $\{\theta_1,\theta_2,\theta_3, \mathbb{R}(c), \mathbb{Im}(c)\} \in \mathcal{N}(0,1)$ are independent random variables drawn from a normal distribution with zero mean and unit variance. The parameter $f$ controls the localization length, serving as an effective disorder strength, {whereas $\lambda$ controls the non-Gaussian content of the circuit.}

The $\ell$-bit operators are related to the physical spin operators by the quasi-local unitary $\hat{U}_\ell$, i.e., $\hat\tau_{k}^{\alpha} = \hat{U}_\ell\hat\sigma_{k}^{\alpha}\hat{U}_\ell^{\dagger}$, with $\alpha\in\{x,y,z\}$. As a consequence, a local spin operator along the $\hat\sigma^z$-direction can be written as
\begin{equation}
    \hat{\sigma}^{z}_k = \sum_{k_1}B_{k_1}\hat{\tau}_{k_1}^{z}  + \sum_{k_1,k_2,\alpha^{\prime}} B_{k_1k_2}^{\alpha^{\prime}\alpha^{\prime}}\hat{\tau}_{k_1}^{\alpha^{\prime}}\hat{\tau}_{k_2}^{\alpha^{\prime}} +\ldots
    \label{Eq:Z_Lbit}
\end{equation}
with $\alpha^{\prime}\in\{x,y,z\}$. The quasi-locality of the $\ell$-bit operators guarantees that the coefficients $B_{k}$ decay exponentially with the distance between the physical site $k$ and the $\ell$-bit indices. Additionally, the operators $\hat{\tau}^{\alpha}$ are chosen such that $[\hat{\sigma}^{z}_k,\sum_{k^{\prime}}\hat\tau_{k^{\prime}}^{z}]=0$. 

We now investigate the dynamics of the Majorana operators and provide a complete derivation of Eq.~\eqref{TD_MBL}.  For $k=1$, the FAF is given by Eq.~\eqref{eq:faf1}.
The two-point Majorana correlators can be expressed, {up to a phase}, as a Pauli string, $\hat\gamma_m\hat\gamma_n = \hat\sigma_{i}^{\alpha}\Big(\prod_{j=i+1}^{l-1} \hat{\sigma}^{z}_{j}\Big)\hat\sigma_{l}^{\alpha^\prime}$, with $\alpha,\alpha^{\prime} \in \{x,y\}$ fixed by the parities of $m$ and $n$, $i=\lceil m/2\rceil$, and $l=\lceil n/2\rceil$. For Majoranas on the same site, this reduces to a local $\hat{\sigma}^z$; for distinct sites, it is a nonlocal string with two nontrivial $(\hat{\sigma}^{x},\hat{\sigma}^y)$ operators connected by a parity string.

It is crucial to first understand the dynamics of the $\ell$-bit operators under the dynamics of Eq.~\eqref{eq:l-bit} and, subsequently, map these results back to spin operators using Eq.~\eqref{Eq:Z_Lbit}. Within the $\ell$-bit framework, we have that $\langle \hat{\tau}_{k}^{z}(t)\rangle = \langle \hat{\tau}_{k}^{z}(0)\rangle$. In contrast, off-diagonal operators exhibit algebraic decay, with distinct exponents depending on whether the corresponding $\ell$-bits are entangled~\cite{serbyn2014quantum}

\begin{equation}\label{Power_law}
    |\langle T^{\alpha\alpha^{\prime}}_{ij}(t)\rangle| \propto \begin{cases}
        1/t^{2\beta}, \quad &t \ll J_0^{-1}e^{|i-j|/\xi}\\
        1/t^{\beta}, \quad &t \gg J_0^{-1}e^{|i-j|/\xi}
    \end{cases}
\end{equation}
where $T^{\alpha\alpha^{\prime}}_{ij}=\hat\tau_{i}^{\alpha}\hat\tau_{j}^{\alpha^\prime}$. Here, $\xi$ is the effective localization length of the $\ell$-bits and $\beta\leq\xi\ln2$. These two different dynamical regimes allow us to write
\begin{align}\label{Eq.2}
    \mathcal{F}_1(\ket{\Psi_t}) =& \left(L - \sum_{k=1}^{L}|\langle\hat{\sigma}^{z}_{k}\rangle|^2_{t}\right) - \sum_{m=1}^{2L}\sum_{n=m+2}^{2L}|\langle\hat\gamma_{m}\hat\gamma_n\rangle|^2_t \,.
\end{align}
The first term on the r.h.s. has a significant overlap with $\hat{\tau}^{z}$'s operators and, therefore, controls the upper bound of $\mathcal{F}_1^{\mathrm{MBL}}$. Nevertheless, the second term on the r.h.s of Eq.~\eqref{Eq:Z_Lbit} guarantees a small power-law decay before reaching its saturation value. The latter term of Eq.~\eqref{Eq.2} exhibits similar power-law behavior, although with a much smaller saturation value. 

The power-law exponent of Eq.~\eqref{Power_law} depends on whether a given pair of $\ell$-bits is entangled. {At time $t$, this is the case for pairs separated by less than $r^{\star}(t) = \xi\ln{(J_0t)}$}. Consequently, the FAF assumes the form

\begin{equation}\label{Eq:Proof_TD}
    \mathcal{F}_1(|\Psi(t)\rangle) = \mathcal{F}_1^{\mathrm{MBL}} - \sum_{r=1}^{\lfloor r^{\star}\rfloor }A_1(r)t^{-\beta} - \sum_{r=\lceil r^{\star}\rceil}A_2(r)t^{-2\beta}
\end{equation}
where the coefficients $A_{1,2}(r)$ encode the exponentially suppressed contributions of distant $\ell$-bit pairs and depend only on the distance between the spins. 

\begin{figure}[t!]
    \centering
    \includegraphics[width=1\linewidth]{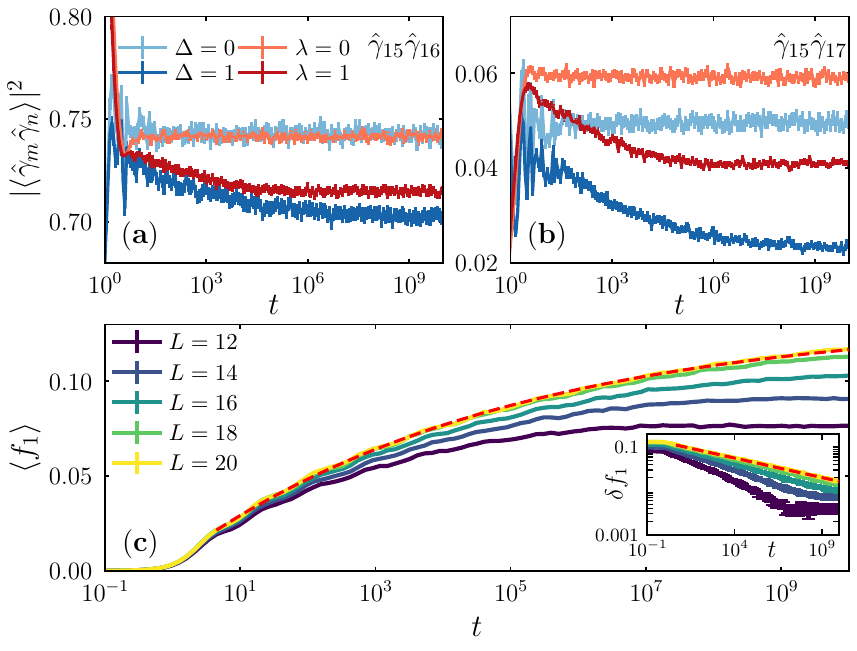}
    \caption{(a-b) Time evolution of Majorana correlators $|\braket{\hat\gamma_m\hat\gamma_n}|^2$ in the MBL regime for the XXZ Hamiltonian (blue) and the $\ell$-bit model (red) for $L=16$ and $W=12$ (for the XXZ model). In the noninteracting case ($\Delta=\lambda=0$, light colors), the correlators rapidly saturate to a constant, reflecting trivial dynamics; in contrast, in the interacting case ($\Delta=\lambda=1$, dark colors), spin dephasing induces a slow power-law decay before saturation. (c) Dynamics of $\langle f_1\rangle$ for the $\ell$-bit Hamiltonian at $\lambda=1$ for several system sizes $L$. The dynamics display a clear power-law behavior (red-dashed line), consistent with Eq.~\eqref{Eq:Proof_TD}. Inset: Time-evolution of $\delta f_1={f}_1^{\mathrm{MBL}} -\langle f_1\rangle$, highlighting the same power-law behavior.}
    \label{Fig:Majoranas_FAF}
\end{figure}

We now investigate the time dynamics of two-point Majorana correlators in the MBL regime, comparing the phenomenological $\ell$-bit description with the microscopic XXZ model. In both systems, we consider an initial N\'eel state and evolve the system using exact methods. As shown in Fig.~\ref{Fig:Majoranas_FAF}(a,b), the dynamics of Majoranas are aligned in both models: In the noninteracting limit ($\Delta=0$ for XXZ and $\lambda=0$ for $\ell$-bit), the correlators $|\braket{\hat\gamma_m\hat\gamma_n}|^2$ rapidly saturate to a constant, with the value depending on whether we consider a local $\hat{\sigma}^z$ operator [panel (a)] or a non-local string [panel (b)]. In contrast, interactions ($\Delta=\lambda=1$, dark colors) induce a slow power-law decay over several decades in time before saturation, a hallmark of MBL dynamics. Since the FAF dynamics are directly inherited from the underlying Majorana correlators, the same power-law behavior is expected to govern their time dependence.

Figure~\ref{Fig:Majoranas_FAF}(c) shows the time evolution of the FAF density $\langle f_1\rangle$ for several system sizes $L$ at fixed $\xi^{\prime}=0.5$ and $\lambda=1$. The FAF exhibits a clear, extensive growth with the system size $L$, and its dynamical behavior is well captured by the phenomenological description of Eq.~\eqref{Eq:Proof_TD}. A fit of the $L=20$ data using Eq.~\eqref{TD_MBL} yields excellent agreement (red dashed-line), with fitted parameters $f_1^{\mathrm{MBL}}\approx 0.133$ and $\beta\approx0.09$. The inset further highlights the power-law relaxation, showing that $\delta f_1={f}_1^{\mathrm{MBL}} -\langle f_1\rangle$ decays algebraically in time.

\clearpage          
\onecolumngrid
\begin{center}
  {\large\bfseries Supplementary material: Fermionic magic resources in disordered quantum spin chains}\\[1.0em]

  Pedro R. Nic\'acio Falc\~ao\,\orcidlink{0000-0001-5893-0460}$^{1,2}$,
  Jakub Zakrzewski\,\orcidlink{0000-0003-0998-9460}$^{2,3}$,
  and Piotr Sierant\,\orcidlink{0000-0001-9219-7274}$^{4}$\\[0.6em]

  {\small\itshape
  $^{1}$Szkoła Doktorska Nauk \'Scis\l{}ych i Przyrodniczych, Uniwersytet Jagiello\'nski, \L{}ojasiewicza 11, PL-30-348 Krak\'ow, Poland\\
  $^{2}$Instytut Fizyki Teoretycznej, Wydzia\l{} Fizyki, Astronomii i Informatyki Stosowanej, Uniwersytet Jagiello\'nski, \L{}ojasiewicza 11, PL-30-348 Krak\'ow, Poland\\
  $^{3}$Mark Kac Complex Systems Research Center, Uniwersytet Jagiello{\'n}ski, Krak{\'o}w, Poland\\
  $^{4}$Barcelona Supercomputing Center, Barcelona 08034, Spain\par}
  \vspace{0.5em}

\end{center}
\vspace{0.5em}
\twocolumngrid

\setcounter{secnumdepth}{2}
\renewcommand{\thesection}{\arabic{section}}
\renewcommand{\thesubsection}{\thesection.\arabic{subsection}}
\renewcommand{\theequation}{S\arabic{equation}}
\renewcommand{\thefigure}{S\arabic{figure}}
\renewcommand{\thetable}{S\arabic{table}}
\renewcommand{\theHequation}{S\arabic{equation}}
\renewcommand{\theHfigure}{S\arabic{figure}}
\renewcommand{\theHtable}{S\arabic{table}}
\setcounter{section}{0}
\setcounter{equation}{0}
\setcounter{figure}{0}
\setcounter{table}{0}

\tableofcontents

\section{Summary}

In this supplementary material, we provide the technical details underlying the results presented in the main text. In Sec.~\ref{Sec:Anderson_Basis}, we expand the interacting Hamiltonian in terms of Majorana modes and show that it splits into four contributions crucial to understanding the non-Gaussian structure of MBL eigenstates. In Sec.~\ref{Sec.Numerics}, we detail the numerical procedure and the statistical uncertainties employed throughout the manuscript. In Sec.~\ref{Perturbative_Analysis}, we develop a perturbative analysis showing that the FAF of \emph{typical} MBL eigenstates is, to leading order, equivalent to sixteen times the probability of creating four quasiparticle excitations above the Anderson eigenstates. In Sec.~\ref{FAF_U1}, we show that, for U(1)-symmetric systems, the FAF is fully determined by the eigenvalues of the single-particle density matrix. We then turn to the non-Gaussianity of cat-like eigenstates. In Sec.~\ref{FAF_cats}, we show that short-range resonances typically involve four adjacent spins, with the FAF acting as a proxy for the number of spins participating close to a perfect resonance, while long-range resonances involve a number of spins that grows with system size. In Sec.~\ref{Mechanism_Cats}, we identify a rare geometric configuration~---~a pair of adjacent resonances~---~that dominates the mean FAF in the strong-disorder limit, where the disorder dependence enters only through the geometric probability of the resonant configuration. In Sec.~\ref{Crossing_Points}, we detail how the FAF crossing points were extracted. In Sec.~\ref{Sec.Time_Evolution}, we present additional analysis of the time evolution of the FAF, discussing the difficulty of numerically distinguishing the predicted power-law saturation from logarithmic growth in experimentally accessible time windows, the FAF's approach to its upper bound, and the robustness of the predicted behavior across different disorder strengths and initial states. Finally, in Sec.~\ref{Sec_FAF_index}, we show that the main conclusions of the manuscript remain qualitatively unchanged for other FAF indices.

\section{Interacting Hamiltonians in the Anderson basis}\label{Sec:Anderson_Basis}

We start by recalling the models studied in the main letter. We consider spin-$1\!/2$ chains of the form $\hat{\mathcal H}=\hat{\mathcal H}_{\rm xx}+\hat{\mathcal H}_{\mathrm{int}}$, where

\begin{equation}
    \hat{\mathcal H}_{\rm xx}
    =\frac{J}{4}\sum_{j=1}^{L}\!\Big(\hat{\sigma}^x_j \hat{\sigma}^x_{j+1}+\hat{\sigma}^y_j \hat{\sigma}^y_{j+1}\Big)
    +\frac{1}{2}\sum_{j=1}^{L} h_j \hat{\sigma}^{z}_j ,
    \label{Hamilt_XX_SP}
\end{equation}
is a non-interacting XX Hamiltonian and we consider $J=1$. Here, $\hat{\sigma}^{\alpha}_j$, with $\alpha \in \{x,y,z\}$, are Pauli operators acting on site $j$, and the fields $h_j$ are drawn independently from a uniform distribution on $[-W,W]$. The interacting contribution is either extensive, $\hat{\mathcal H}_{\mathrm{ext}}=\tfrac{\Delta}{4}\sum_{j=1}^{L}\hat{\sigma}^z_j\hat{\sigma}^z_{j+1}$, for the XXZ Hamiltonian, or confined to a single bond, $\hat{\mathcal H}_{\mathrm{conf}}=\tfrac{\Delta}{4}\,\hat{\sigma}^z_{L/2}
\hat{\sigma}^z_{L/2+1}$, for the impurity model. The XX Hamiltonian $\hat{\mathcal H}_{\rm xx}$ admits a free-fermion representation upon a Jordan--Wigner transformation [see Eq.~\eqref{eq:WJ}],

\begin{equation}
    \hat{\mathcal{H}}_\mathrm{xx}=\tfrac{i}{4}\sum_{j=1}^{L}\eta_j\big(\hat\gamma_{2j-1}\hat\gamma_{2j+2} - \hat\gamma_{2j}\hat\gamma_{2j+1}\big) - \tfrac{i}{2} \sum_{j=1}^{L}h_j\hat\gamma_{2j-1}\hat\gamma_{2j} \,,
\end{equation}
with $\eta_j=1$ for $j\neq L$ and $\eta_L=\pm1$ fixed by the fermionic parity sector. We can define a transformation such that the XX Hamiltonian can be written in its normal-mode decomposition, $\hat{\mathcal{H}}_\mathrm{xx}=\sum_{\alpha=1}^{L}\varepsilon_\alpha\hat{Q}_{\alpha}$, with $\hat{Q}_\alpha=\tfrac{i}{2}\tilde{\gamma}_{2\alpha-1}\tilde{\gamma}_{2\alpha}$ (see the main text for definitions). The operators $\hat{Q}_{\alpha}$ are conserved quantities that define the Anderson local integrals of motion (ALIOMs).

To express the interacting disordered Hamiltonians in the ALIOM basis, we introduce new fermionic operators $\hat{f}_{\alpha}=\tfrac{1}{2}(\tilde\gamma_{2\alpha-1} + i\tilde\gamma_{2\alpha})$, satisfying $\{\hat{f}_\alpha,\hat{f}_{\beta}^{\dagger}\}=\delta_{\alpha\beta}$ and $\hat{Q}_{\alpha} = \hat{f}_{\alpha}^{\dagger}\hat{f}_{\alpha}-\tfrac{1}{2}$. In terms of these modes, the interacting contribution of the XXZ Hamiltonian takes the form $\hat{\mathcal{H}}_\mathrm{ext} = \sum_{k,l,m,n} U_\mathrm{ext}^{k,l,m,n} \hat{f}_k^\dagger \hat{f}_l \, \hat{f}_m^\dagger \hat{f}_n$ and, as shown in  Ref.~\cite{Laflorencie20}, $\hat{\mathcal{H}}_\mathrm{ext}$ can be decomposed into four different contributions,

\begin{align}\label{Interacting_Part}
    \hat{\mathcal{H}}_\mathrm{ext} &= \sum_{k=1}^{L} \mathcal{U}_k^{(1)}\hat{Q}_k + \sum_{k\neq l} \mathcal{U}_{k,l}^{(2)}\hat{Q}_k\hat{Q}_l \\
     &+\sum_{k\neq l \neq m} \mathcal{U}_{k,l,m}^{(3)} \hat{Q}_k \hat{f}^{\dagger}_{l}\hat{f}_{m} + \sum_{k\neq l \neq m \neq n} \mathcal{U}_{k,l,m,n}^{(4)} \hat{f}_{k}^{\dagger}\hat{f}_{l}\hat{f}_{m}^{\dagger}\hat{f}_{n}\nonumber 
\end{align}
where the first line of Eq.~\eqref{Interacting_Part} commutes with the ALIOMs and only renormalizes the many-body energies, while the second line does not commute with them and constitutes the genuine perturbation that dresses the eigenstates. Among these terms, the $\mathcal{U}^{(3)}$ term generates single particle-hole excitations and, at leading order, dresses the eigenstate within the Gaussian manifold. In contrast, the $\mathcal{U}^{(4)}$ term drives the eigenstate out of the Gaussian manifold already at first order.

In the strong disorder regime, the off-diagonal terms of Eq.~\eqref{Interacting_Part} decay exponentially with the distance between the involved orbitals and are therefore dominated by maximally overlapping configurations. The leading contributions act on successive sites and scale as
\begin{align}
    & \mathcal{U}_{k,k+1,k+2}^{(3)} \propto \Delta/W \\
    & \mathcal{U}_{k,k+1,k+2,k+3}^{(4)} \propto \Delta/W^{2}
\end{align}

Since these terms do not commute with the ALIOMs, they constitute, in the perturbative picture, the source of the generation of fermionic magic resources in disordered systems. At small $\Delta$, the dominant non-Gaussian channel is the first-order $\mathcal{U}^{(4)}$ term, while the second-order process in $\mathcal{U}^{(3)}$~---~two successive single-particle hops needed to populate the same $k=4$ sector~---~contributes with a much smaller amplitude at strong disorder.

A similar representation can be obtained for the interacting term in the impurity model, $\hat{\mathcal{H}}_\mathrm{conf}$. However, the presence of confined interactions in the Hamiltonian limits the expansion of the interacting terms only to the sites near $L/2$, resulting in an \emph{area-law} scaling of the FAF.

\section{Numerical methods and statistical uncertainty}~\label{Sec.Numerics}

In this section, we provide additional details on the numerical procedures used throughout the main text. We consider the fermionic antiflatness (FAF) as a measure of non-Gaussianity~\cite{Sierant26faf}, 
\begin{equation}
    \mathcal{F}_k = L - \tfrac{1}{2}\mathrm{tr}\left[\left(M^TM\right)^k\right]\,,
\end{equation}
where $M$ is the covariance matrix, with elements $M_{mn} = -\tfrac{i}{2}\braket{[\hat{\gamma}_m,\hat\gamma_n]}$, $k\geq1$ is an integer, and $\{\nu_\alpha\}$ is the Williamson eigenspectrum~\cite{Williamson36} of $M$, related to $\mathcal{F}_k$ via $\mathcal{F}_k = \sum_{\alpha=1}^{L}(1-\nu_\alpha^{2k})$, with $\nu_\alpha^2=1$ for every $\alpha$ if and only if the state is a pure Gaussian state. Unless otherwise stated, we set $k=1$, the most sensitive choice for partial mode occupations.

We employ exact diagonalization, using full diagonalization for $L\leq14$ and the POLFED algorithm~\cite{Sierant20p} for $L\geq16$. For each disorder realization, we target $N_{\rm ev}=\min\{100,\mathcal{N}/20\}$ eigenstates closest to the energy density $\epsilon=0.5$, where $\epsilon = (E-E_{\rm min})/(E_{\rm max}-E_{\rm min})$ and $\mathcal{N}$ is the Hilbert space dimension. This choice provides a uniform sample of eigenstates from the bulk of the spectrum, where the density of states is approximately flat. The average value $\langle \mathcal{F}_k\rangle$ is calculated in two steps: for each disorder realization, we first average $\mathcal{F}_k$ over the $N_{\rm ev}$ eigenstates; subsequently, we average over all $N_{\rm dis}$ different disorder realizations. The median, in contrast, is taken directly over the pooled $N_{\rm ev}\times N_{\rm dis}$ eigenstates.

We estimate the statistical error by computing the standard error of the mean,
\begin{equation}
    \sigma_{X} = \sqrt{\frac{\braket{(X_S - \braket{X})^2}}{N_{\rm dis}}} \,,
\end{equation}
where $X\equiv\mathcal{F}_k$ and $X_S$ denote the eigenstate-averaged value of $X$ for disorder realization $S$, as defined above; $\braket{\cdot}$ denotes the average over the $N_{\rm dis}$ realizations. The $1/\sqrt{N_{\rm dis}}$ scaling assumes that $X_S$ is statistically independent across disorder samples, so the variance of the mean decreases with the number of independent realizations rather than any residual correlation between them.

The number of disorder realizations $N_{\rm dis}$ is chosen to balance statistical accuracy against the increasing cost of exact diagonalization with system size. For $L\leq12$, we consider at least $10^5$ disorder realizations; for $L=14$--$18$, at least $10^4$, increased to ${O}(10^5)$ in the vicinity of the crossing point $W^\star_{\langle f_1\rangle}$ (Sec.~\ref{Crossing_Points}) to determine it accurately; and for $L=20$, where the Hilbert-space dimension $\mathcal{N}=\binom{20}{10}=184756$ significantly limits statistics, at least $10^3$ realizations, increased to up to $3\times10^4$ near the crossing point. These realization counts apply throughout the manuscript and Supplementary Material unless stated otherwise.

\section{Perturbative analysis and typical scaling of the FAF in the MBL regime}\label{Perturbative_Analysis}

In the main letter, we present Eq.~4 as an analytical solution for the FAF deep in the MBL regime, showing that the FAF of typical MBL eigenstates reduces to sixteen times the population of the four-quasiparticle sector. In this section, we present a detailed derivation of this expression and additional numerical data that support this result.

Let us assume the existence of a unitary operator $\hat{U}_M = e^{\hat{S}}$, with $\hat{S}$ being an anti-Hermitian operator that maps an eigenstate of a free-fermionic $\hat{\mathcal{H}}_{\rm xx}$ Hamiltonian onto an eigenstate of $\hat{\mathcal{H}}_{\rm xxz}$, i.e., $\ket{\Psi^{\rm MBL}}=e^{\hat{S}}\ket{\Psi^{\rm And}}$. For brevity, we write $\ket{0}\equiv \ket{\Psi^{\rm And}}$ and $\ket{\Psi}\equiv \ket{\Psi^{\rm MBL}}$, taking $\lVert\hat{S}\rVert\ll1$ throughout, as such an operator must be quasi-local. The covariance matrix of the interacting eigenstate $\ket{\Psi}$ then reads
\begin{equation}
    M_{mn} = -\tfrac{i}{2}\bra{0}\left[\hat{\gamma}_{m}^{\prime}, \hat{\gamma}_{n}^{\prime}\right]\ket{0}\, ,
    \label{Eq.New_Cov}
\end{equation}
where $\hat{\gamma}_{m}^{\prime}=\hat{U}_M^{\dagger}\hat{\gamma}_{m}\hat{U}_{M}=e^{-\hat{S}}\hat{\gamma}_{m}e^{\hat{S}}$ are the rotated Majoranas. Applying the Baker-Campbell-Hausdorff (BCH) expansion, we can write the covariance matrix as
\begin{equation}
    M = M^0 + \sum_{l=1}^{\infty}\delta M^{(l)} \,,
\end{equation}
where $M^{0}$ is the covariance matrix of the unperturbed eigenstate $\ket{0}$, while $\delta M^{(l)}$ denotes the ${O}(\hat{S}^{(l)})$ correction induced by the interactions. Inserting this expansion into the definition of the FAF (Eq.~\eqref{FAF}) yields 

\begin{equation}
    \mathcal{F}_{1}\left(\ket{\Psi}\right) = \sum_{l=1}^{2}\operatorname{tr}\left(M^{0}\delta M^{(l)} \right)  - \tfrac{1}{2}\lVert \delta M^{(1)}\rVert_{F}^{2} + \cdots \,,
\end{equation}
with the ellipsis denoting terms of order ${O}(\hat{S}^3)$ and higher. Such terms are negligible for \emph{typical} MBL eigenstates, for which $\lVert \hat{S}\rVert \ll 1$.

We now show that the leading contribution, $T = \operatorname{tr}\left(M^{0}\delta M^{(1)} \right)$, vanishes. A straightforward calculation shows that this term can be written as
\begin{equation}
    T = 4\left\langle\left[\hat{K}_0,\hat{S}\right]\right\rangle_0 \,,
    \label{Eq.Leading_FAF}
\end{equation}
where $\hat{K}_{0} = (i/4)\sum_{mn}M_{mn}^{0}\hat{\gamma}_m\hat{\gamma}_n$ is a parent Hamiltonian and $\left\langle \left(\cdots\right)\right\rangle_0 \equiv \bra{0} \left(\cdots\right)\ket{0}$. The covariance matrix can be brought to its canonical form $M^0 = G\left(\bigoplus_{\alpha=1}^{L}is_{\alpha}\hat{\sigma}^{y}\right)G^{T}$, where $G$ is an orthogonal matrix and $s_\alpha=\pm1$ are the Williamson eigenvalues~\cite{Williamson36}. Introducing new rotated Majorana operators as $\tilde\gamma_{a} = \sum_{m}G_{ma}\hat{\gamma}_m$, the parent Hamiltonian takes the diagonal form
\begin{equation}
    \hat{K}_0 = \frac{1}{2}\sum_{\alpha=1}^{L} s_{\alpha}\hat{L}_{\alpha}\,, \qquad \hat{L}_{\alpha}= i\tilde{\gamma}_{2\alpha-1}\tilde{\gamma}_{2\alpha}\,.
    \label{Eq.Parent}
\end{equation}

Each $\hat{L}_{\alpha}$ is Hermitian and squares to the identity $\hat{L}_{\alpha}^{2} = \hat{\mathds{1}}$, and therefore has eigenvalues $\pm1$. The state $\ket{0}$ is a simultaneous eigenstate of all $\hat{L}_{\alpha}$, with $\hat L_\alpha\ket{0}=-s_\alpha\ket{0}$, so that
$\hat K_0\ket{0}=E_0\ket{0}$, with $E_0=-L/2$. Since $\ket{0}$ is an eigenstate of the parent Hamiltonian $\hat{K}_0$, the expectation value of the commutator of Eq.~\eqref{Eq.Leading_FAF} vanishes, giving $T=0$. The leading $O(\hat S)$ contribution to the FAF is thus identically zero; hence, the dominant term in the perturbative analysis of the FAF is ${O}(\hat{S}^2)$.

We now turn to the leading nonvanishing contribution to the FAF. The second-order correction to the covariance matrix reads
\begin{equation}
    \delta M_{mn}^{(2)} = \tfrac{1}{2}\left\langle \mathrm{ad}_S^{2}\hat{\Gamma}_{mn} \right\rangle_0 \,,\quad \hat{\Gamma}_{mn}=-\tfrac{i}{2}\left[\hat{\gamma}_m,\hat{\gamma}_n\right]\,,
    \label{Eq.dM2_1}
\end{equation}
where $\mathrm{ad}_S\left(\cdot\right)=[\hat{S}, \cdot]$ denotes the adjoint action of $\hat{S}$. A straightforward calculation allows us to write $P = \operatorname{tr}\left(M^0\delta M^{(2)}\right)= 2\left\langle\mathrm{ad}_{S}^{2}\hat{K}_0\right\rangle_0$ and, since $\ket{0}$ is an eigenstate of $\hat{K}_0$ with eigenvalue $E_0$, this can be recast as
\begin{equation}
    P = 4\bra{\Phi}\left(\hat{K}_0 - E_0\right)\ket{\Phi} \geq0\,,
\end{equation}
with $\ket{\Phi}=\hat{S}\ket{0}$. We now introduce occupation modes relative to the Anderson eigenstate,

\begin{equation}
    \hat{n}=\sum_{\alpha=1}^{L}\hat{n}_{\alpha} \,; \qquad \hat{n}_{\alpha} = \tfrac{1}{2}\left(1 + s_{\alpha}\hat{L}_{\alpha}\right) \,,
    \label{Eq.Number_Operator}
\end{equation}
where each $\hat{n}_{\alpha}$ has eigenvalues in $\{0,1\}$ and $\hat{n}=\hat{K}_0 -E_0\succeq0$. These occupation modes are naturally expressed through the Bogoliubov modes, 
\begin{equation}
    \hat{b}_{\alpha} = \tfrac{1}{2}\left(\tilde\gamma_{2\alpha-1} + is_{\alpha}\tilde{\gamma}_{2\alpha}\right) \,,
    \label{Eq.Bogoliubov}
\end{equation}
which satisfies the canonical anticommutation relation $\{\hat{b}_{\alpha}, \hat{b}_{\beta}^{\dagger}\}=\delta_{\alpha\beta}$ and obeys $\hat{n}_{\alpha}=\hat{b}_{\alpha}^{\dagger}\hat{b}_{\alpha}$. Since every $\hat{n}_{\alpha}$ annihilates $\ket{0}$, the Anderson eigenstate plays the role of a Bogoliubov vacuum, and $\hat{n}=\sum_{\alpha=1}^{L}\hat{n}_{\alpha}$ counts the number of quasiparticle excitations above it. Decomposing $P$ into different excitation sectors and noting that a physical generator $\hat{S}$ has even fermionic parity, so that only even sectors contribute, we obtain

\begin{equation}
    P = 4\sum_{k\geq2} k\,\lVert\hat\Pi_{k}\hat{S}\ket{0}\rVert^{2}\,,
    \label{Eq.P}
\end{equation}
where $\hat\Pi_{k} = \ket{k}\bra{k}$ projects onto the $k$-th excitation sector. The $k=2$ sector is reachable by a Gaussian unitary; since the FAF is invariant under such transformations, the contribution of this sector must be canceled by the term $(1/2)\lVert\delta M^{(1)}\rVert_{F}^{2}$. 

We now show that this is indeed the case. Starting from the first-order correction of the covariance matrix and using that $\hat{S}$ is anti-Hermitian, the matrix elements of $\delta M^{(1)}$ can be written as

\begin{equation}
    \delta M^{(1)}_{mn} = \left\langle\left[\hat{\Gamma}_{mn},\hat{S}\right]
    \right\rangle_0 = 2\,\mathrm{Re}\bra{0}\hat{\Gamma}_{mn}\ket{\Phi}\,,
\end{equation}
where we used the Hermiticity property of $\hat{\Gamma}_{mn}$. Since $\hat{\Gamma}_{mn}$ is quadratic in Majoranas, it changes the excitation number by $0$ or $\pm 2$; inserting the completeness relation $\sum_k \ket{k}\bra{k}=\hat{\mathds{1}}$, the first-order correction of the covariance matrix reads

\begin{equation}
    \delta M^{(1)}_{mn} = 2\,\mathrm{Re}\left(\bra{0}\hat{\Gamma}_{mn}\hat\Pi_0
    \ket{\Phi} + \bra{0}\hat{\Gamma}_{mn}\hat\Pi_2\ket{\Phi}\right)\,.
    \label{Eq.DM1_P1}
\end{equation}

The first term, however, vanishes since $\bra{0}\hat{\Gamma}_{mn}\hat\Pi_0\ket{\Phi}=M_{mn}^{0}\langle \hat{S}\rangle_0=0$, as $M_{mn}^{0}$ is real while $\langle \hat{S}\rangle_0$ is purely imaginary; thus, only the second term on the r.h.s of Eq.~\eqref{Eq.DM1_P1} remains. Defining $\ket{\Phi_2}\equiv\hat{\Pi}_2\ket{\Phi}=\sum_{\alpha<\beta}\phi_{\alpha\beta}\hat{b}_{\alpha}^{\dagger}\hat{b}_{\beta}^{\dagger}\ket{0}$ as the projected state and $z_{mn}\equiv\bra{0}
\hat{\Gamma}_{mn}\ket{\Phi_2}$, we have $\delta M^{(1)}_{mn}=z_{mn}+z_{mn}^{*}$,
so that
\begin{equation}
    \lVert\delta M^{(1)}\rVert_F^2 = 2\sum_{mn}|z_{mn}|^2
    + 2\,\mathrm{Re}\sum_{mn}z_{mn}^2\,.
    \label{Eq.AB}
\end{equation}

Both sums are frame invariant. Writing $\hat{\gamma}_m=\sum_a G_{ma}\tilde\gamma_a$ gives $z_{mn}=\sum_{ab}G_{ma}G_{nb}\tilde{z}_{ab}$ with $\tilde{z}_{ab} =\bra{0}\tilde\Gamma_{ab}\ket{\Phi_2}$, and orthogonality $\sum_m G_{ma}G_{mc}=\delta_{ac}$ yields $\sum_{mn}|z_{mn}|^2=\sum_{ab} |\tilde{z}_{ab}|^2$ and $\sum_{mn}z_{mn}^2=\sum_{ab}\tilde{z}_{ab}^2$. Expressing the rotated Majoranas in terms of the Bogoliubov modes, $\tilde\gamma_a =u_a\hat{b}_\alpha+v_a\hat{b}^\dagger_\alpha$ with

\begin{equation}
    (u_a,v_a)=\begin{cases}(1,1)\,, & a=2\alpha-1\,,\\[2pt]
    (-is_\alpha,\,is_\alpha)\,, & a=2\alpha\,,\end{cases}
\end{equation}
the coefficients satisfy $|u_{a}|^{2}=1$ and $\sum_{a\in\alpha} u_{a}^{2} = 1 + (-is_{\alpha})^2 = 0$. If $a$ and $b$ lie in the same block, $\tilde\Gamma_{ab}=-\hat{L}_\alpha$ conserves the excitation number and gives no overlap with $\ket{\Phi_2}$. For $a$ and $b$ in different blocks ($\alpha\neq\beta$),
\begin{equation}
    \tilde{z}_{ab}=-i\,u_a u_b\,w_{\alpha\beta}\,,
\end{equation}
with $w_{\alpha\beta}=\bra{0}\hat{b}_\alpha\hat{b}_\beta\ket{\Phi_2}=-\phi_{\alpha\beta}$ being antisymmetric, such that $\sum_{\alpha\neq\beta} |w_{\alpha\beta}|^2=2\sum_{\alpha<\beta}|\phi_{\alpha\beta}|^2=2\lVert\Phi_2\rVert^2$. The first term on the r.h.s of Eq.~\eqref{Eq.AB} reads

\begin{equation}
    2\sum_{mn}|z_{mn}|^2=  8\sum_{\alpha\neq\beta}|w_{\alpha\beta}|^2
    =  16\lVert\hat\Pi_2\Phi\rVert^2\,,
\end{equation}
where we used the rotation-invariant property of $z_{mn}$, i.e., $\sum_{mn}|z_{mn}|^2=\sum_{ab}|\tilde{z}_{ab}|^2$. 

The second term on the r.h.s of Eq.~\eqref{Eq.AB} vanishes. Since $z_{mn}^{2} = -u_a^2u_b^2w_{\alpha\beta}^2$, this term can be expressed as 

\begin{equation}
    \sum_{ab}\tilde{z}_{ab}^2 = -\sum_{\alpha\neq\beta}w_{\alpha\beta}^2
    \left(\sum_{a\in\alpha}u_a^2\right)\!\left(\sum_{b\in\beta}u_b^2\right)=0\,,
\end{equation}
but the sum over each block vanishes, showing that $\sum_{mn}{z}_{mn}^2=0$. Inserting both expressions in Eq.~\eqref{Eq.AB} yields

\begin{equation}
    \tfrac{1}{2}\lVert \delta M^{(1)}\rVert_{F}^{2} = 8\lVert \hat{\Pi}_2\hat{S}\ket{0}\rVert^{2}\,,
\end{equation}
which is precisely the $k=2$ term in Eq.~\eqref{Eq.P}. The two terms cancel each other, removing the Gaussian-reachable sector and leaving

\begin{equation}
    \mathcal{F}_1(\ket{\Psi}) = 4\sum_{k\geq4}\,k\lVert\hat\Pi_{k}\hat{S}\ket{0}\rVert^{2} + {O}(\lVert\hat{S}\rVert^3)\,,
\end{equation}
showing the validity of Eq.~\eqref{Eq:F1_Scaling}. 

The structure of the interacting part of the XXZ Hamiltonian, presented in Eq.~\eqref{Interacting_Part}, allows us to obtain a deeper understanding of the behavior of the FAF. Since $\hat{S}\sim \hat{\mathcal{H}}_{\rm int}^{\perp}/\lVert\hat{\mathcal{H}}_{\rm xx}\rVert$ (using $\lVert\hat{\mathcal{H}}_{\rm xx}\rVert\sim W$ as the characteristic energy denominator), we can decompose

\begin{equation}
    \hat{S} = \hat{S}_{2} + \hat{S}_{4}\,,
\end{equation}
where $\hat{S}_{2}$ ($\hat{S}_{4}$) accounts for the third (fourth) term presented in Eq.~\eqref{Interacting_Part}. The structure of the third term, however, can only generate $k=2$ excitations on the Anderson eigenstate $\ket{0}$ and, as we just showed, such terms are ``invisible'' in the FAF, as they are generated by Gaussian operators. The fourth term of Eq.~\eqref{Interacting_Part}, on the other hand, can generate up to $k=4$ excitations, showing that in the MBL regime of the XXZ Hamiltonian, 

\begin{equation}\label{Eq.Pertub_FAF_exact}
    \mathcal{F}_1(\ket{\Psi}) \simeq 16\lVert\hat{\Pi}_4\hat{S}_4\ket{0}\rVert^2\,,
\end{equation}
yielding $\mathcal{F}_1(\ket{\Psi})\sim \Delta^2/W^{6}$ in the MBL regime, as $\hat{S}_4\propto \Delta/W^{3}$. It is worth noting that such an expression holds only for \emph{typical} MBL eigenstates, where $\hat{U}_{M}$ is a quasi-local unitary for which the BCH expansion is controlled.

\begin{figure}[t!]
    \centering
    \includegraphics[width=0.85\linewidth]{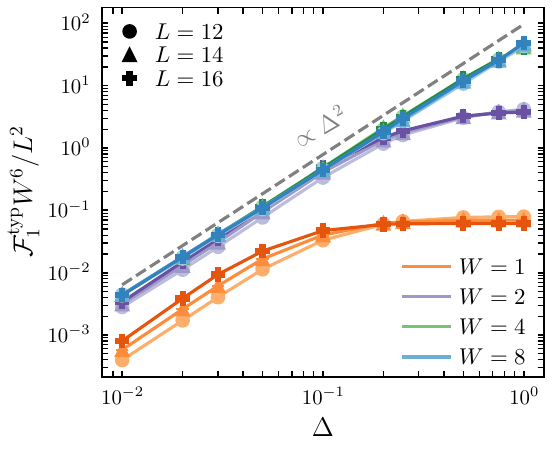}
    \caption{Rescaled typical FAF $\mathcal{F}_1^{\rm typ}W^{6}/L^2$ versus interaction strength $\Delta$ in the XXZ chain ($N_{\rm int}=L$), for $L=12,14,16$ and several disorder strengths. At strong disorder, the curve collapses onto a common $\Delta^2$ scaling (dashed line), confirming $\mathcal{F}_1^{\rm typ}\propto \Delta^2N_{\rm int}^2/W^6$ of Eq.~\eqref{Eq.Pertub_FAF_exact}. Deviations at $W=1,2$ and large $\Delta$ mark the breakdown of the perturbative description.}
    \label{FAF_typ_Delta}
\end{figure}

In Fig.~5 of the main letter, we show the scaling of the \emph{typical} FAF as a function of $W$ for different system sizes, confirming the scaling law predicted by Eq.~\eqref{Eq.Pertub_FAF_exact}. We further substantiate the validity of this scaling by showing in Fig.~\ref{FAF_typ_Delta} the typical FAF as a function of the interaction strength $\Delta$ for different system sizes $L$ and disorder strengths $W$. We rescale the FAF, $\mathcal{F}_1^{\rm typ}\to\mathcal{F}_1^{\rm typ}W^{6}/L^2$, to isolate the dependence on $\Delta$. At weak disorder and sufficiently large $\Delta$, the scaling predicted by Eq.~\eqref{Eq.Pertub_FAF_exact} breaks down, in close agreement with the interaction-driven instabilities recently reported in Ref.~\cite{colbois24interaction}. At strong disorder, however, the perturbative dressing still holds, and the curves collapse onto a single curve that scales $\propto \Delta^2$, as shown by the gray dashed line, numerically demonstrating the validity of Eq.~\eqref{Eq.Pertub_FAF_exact} and our analytical approach.

\section{Fermionic antiflatness in U(1) symmetric systems}\label{FAF_U1}

Here, we show that the FAF can be written as a simple function of the natural-orbital occupations $n_l$. We derive this relation explicitly from the covariance matrix, showing that U(1) symmetry reduces $\mathcal{F}_1$ to the spectrum of the one-particle density matrix. The perspective of one-particle density matrix occupation on MBL was considered earlier in Refs.~\cite{bera2015many, bera2017one, Villalonga18, orito2021multifractality}.

A Jordan-Wigner transformation~\cite{Jordan28} maps the spin chain onto spinless fermions $\hat{c}_i$, with $\{c_i,c_j^\dagger\}=\delta_{ij}$, in terms of which the Majorana operators~[see Eq.~(2) of the main letter] read 

\begin{equation}
    \hat{\gamma}_{2j-1} = \hat{c}_j + \hat{c}_j^{\dagger}\,, \qquad \hat{\gamma}_{2j}= i(\hat{c}_j^{\dagger} - \hat{c}_j)\,,
\end{equation}
so that $\{\hat{\gamma}_m,\hat{\gamma}_n\}=2\delta_{mn}$. Since $\ket{\Psi}$ has a definite particle number, we have $\langle c_ic_j\rangle = \langle c_{i}^{\dagger}c_{j}^{\dagger}\rangle=0$ for any $i,j\in\{1,\ldots,L\}$, whereas $\langle c_i^{\dagger}c_j\rangle$ generically nonzero. The one-particle density matrix $\rho_{ij} = \langle c_i^{\dagger}c_j\rangle$ is Hermitian, and its eigenvectors define the natural orbitals $\ket{\phi_l}$,

\begin{equation}
    {\rho}\ket{\phi_l} = n_l\ket{\phi_l}\,,
\end{equation}
with $n_l\in[0,1]$ being the occupations in the natural basis, with $\sum_{l}n_l = L/2$ at half-filling. The covariance matrix $M_{mn}=-\tfrac{i}{2}\bra{\Psi}[\hat\gamma_m,\hat\gamma_n]\ket{\Psi}$, with $m,n =1,\ldots,2L$, is now entirely determined by the one-particle density matrix $\rho$.

To show this, let $G = 2\rho -\mathbb{1}$ and write $G=A+iB$, with $A = \mathrm{Re}(G)$ and $B = \mathrm{Im}(G)$ as the real and imaginary part of $G$. After ordering the Majoranas, the covariance matrix can be expressed as

\begin{equation}
    M = \begin{pmatrix}
        B & -A \\
        A & B
    \end{pmatrix}\,,
\end{equation}
from where $\mathrm{tr}(M^T M)= 2\,\mathrm{tr}(G^2)\,$. Recalling that $\mathcal{F}_1 =L - \tfrac{1}{2}\mathrm{tr}(M^TM)$ [Eq.~3 of the main letter], and using $G^2 = 4\rho^2 - 4\rho +\mathbb{1}$ together with $\mathrm{tr}(\mathbb{1}_L) = L$, we obtain

\begin{equation}
    \mathcal{F}_1 = 4\mathrm{tr}\left( \rho - \rho^2\right)\, =4\sum_{l=1}^{L}n_l(1-n_l)\,,
\end{equation}
which depends only on the spectrum of $\rho$ and vanishes if and only if $n_l\in\{0,1\}$ for every $l$, i.e., if and only if $\ket{\Psi}$ is a Gaussian state. This measure is equivalent to the Tsallis-$\alpha$ entropy at $\alpha=2$, a recently introduced measure that has been proven to be a strong monotone of fermionic magic resources~\cite{sonya2026computable, turkeshi2026williamson}.

\section{Relation between cat-like MBL eigenstates and the fermionic antiflatness}\label{FAF_cats}

In the main letter, we show that for a perfect cat-like eigenstate $\ket{\Psi_\mathrm{cat}^{\pm}}$~[see Eq.~\eqref{MBL_Cats}], the fermionic antiflatness isolates the contribution of the cat component and provides a proxy for the number of resonant spins. In this section, we expand on the analysis of the cat-like eigenstates. We begin by exploring a simple two-level description for the origin of $\ket{\Psi_\mathrm{cat}^{\pm}}$, showing that the FAF of these states follows a Lorentzian behavior as a function of the detuning, with $\mathcal{F}_1$ counting the number of resonant spins only near a perfect resonance. We show the probability distributions of $\mathcal{F}_1$ for different disorder strengths, numerically confirming that, in the MBL regime, the minimal resonance involves four spins rather than two, with $P(\mathcal{F}_1)$ peaking at $\mathcal{F}_1\approx 4$ for different disorder strengths and system sizes. We then explore FAF of long-range cats, showing that these rare eigenstates substantially increase non-Gaussianity deep in the MBL regime.

\subsection{Two-level description and minimal resonant cluster}\label{Two_Level}

We begin by analyzing the simplest two-level description that produces cat-like eigenstates. Consider two Anderson eigenstates, $\ket{a}$ and $\ket{b}$, which differ across $n$ sites. These configurations hybridize via a matrix element $V$ and are described by the following $2\times2$ Hamiltonian,

\begin{equation}
\hat{\mathcal{H}}_{2\times2} = \begin{pmatrix}
       E_a & V\\
       V & E_b
    \end{pmatrix}\,.
\end{equation}

The eigenstates are $\ket{\Psi^{+}}=\cos\theta\ket{a} + \sin\theta\ket{b}$ and $\ket{\Psi^{-}}=-\sin\theta\ket{a} + \cos\theta\ket{b}$, where $\tan 2\theta = 2V/\delta E$ and $\delta E \equiv E_a - E_b$ denote the energy detuning. In U(1) symmetric systems, such as the XXZ Hamiltonian, the FAF can be expressed as a simple function of the natural orbital occupations,

\begin{equation*}
    \mathcal{F}_1\left(\ket{\Psi}\right) = 4\sum_{l=1}^{L}n_l(1-n_l)\,,
\end{equation*}
where $n_{l}$ denotes the occupation of the $l$-th natural orbital. Using the eigenstates $\ket{\Psi^{\pm}}$ and the fact that $\ket{a}$ and $\ket{b}$ differ only across $n$ sites, it follows that the FAF exhibits Lorentzian behavior,

\begin{equation}
    \mathcal{F}_1(\ket{\Psi^{\pm}}) = \frac{n}{1 + x^2}, \qquad n\geq 4, 
    \label{Lorentzian_FAF}
\end{equation}
where $x = \delta E/2V$. Thus, the FAF saturates at a number $n$ of hybridized spins at exact resonance, and it is bounded by it away from the resonance. The restriction $n\geq 4$ is necessary because, although cat states differing by $n=2$ sites are the most common type of resonance, these configurations differ by a single fermion move, i.e., a Gaussian unitary, under which $\mathcal{F}_1$ is invariant; such states therefore yield $\mathcal{F}_1 = 0$ for all $\theta$. The minimal cat eigenstate with $\mathcal{F}_1 \neq 0$ involves $n=4$ sites, connected by at least two fermion moves via the interacting Hamiltonian $\hat{\mathcal{H}}_{\rm ext}$.

\subsection{Distribution of the FAF and minimal resonances }

\begin{figure}[t!]
    \centering
    \includegraphics[width=1\linewidth]{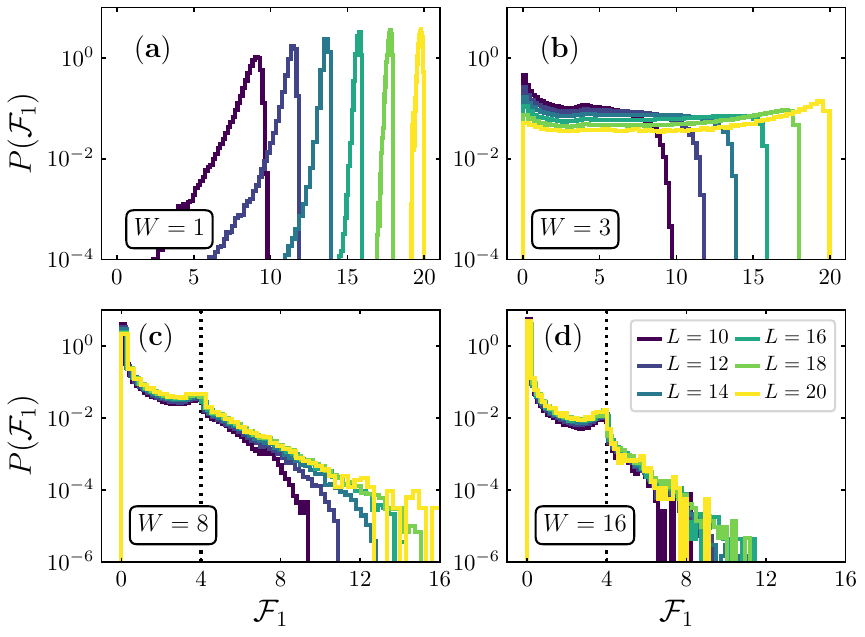}
    \caption{Probability distribution $\mathcal{F}_1$ for different disorder strengths $W$ and system sizes $L$. For $W=1$ [panel (a)], the distribution is approximately Gaussian, with its modes shifting linearly with $L$. For $W=3$ [panel (b)], the distribution becomes broad, a hallmark of the ETH-MBL crossover. In the MBL regime ($W>5$) [panels (c-d)], it develops a secondary peak near $\mathcal{F}_1\approx 4$, and its tail decays exponentially, regardless of the system size. }
    \label{Fig:Distribution_F1}
\end{figure}

We now investigate the probability distribution of $\mathcal{F}_1$ for midspectrum eigenstates of the XXZ Hamiltonian for different disorder strengths $W$ and system sizes $L$. In the simulations, we set $\Delta=1$ and average the results over at least $10^3$ disorder realizations. In the ergodic regime ($W=1$), the position of the mode of ${P}(\mathcal{F}_1)$ shifts linearly with $L$, as illustrated in Fig.~\ref{Fig:Distribution_F1}(a), highlighting the extensive nature of the FAF for ergodic eigenstates. Increasing the disorder strength leads to a qualitative change in the distribution, with the emergence of a long tail extending to extremely low values of $\mathcal{F}_1$. The relative weight of such eigenstates grows with $W$, resulting in a broad distribution at $W=3$, as shown in Fig.~\ref{Fig:Distribution_F1}(b). Similar broad distributions have already been reported for several observables~\cite{Luitz16longTail,Sierant19Level,Luitz20Abs,Gray20Sgap}, and they are a hallmark of the ETH-MBL crossover.

At strong disorder $W=8$, the distribution changes markedly. The primary peak shifts towards $\mathcal{F}_1\approx0$, while a secondary peak emerges at $\mathcal{F}_1\approx4$, independent of the system size. In addition, the tail of the distribution, extending towards $\mathcal{F}_1>4$, decays exponentially, as seen in Fig.~\ref{Fig:Distribution_F1}(c). The tail is strongly suppressed upon the further increase of $W$, and the peak at $\mathcal{F}_1\approx4$ becomes more pronounced, as illustrated in Fig.~\ref{Fig:Distribution_F1}(d). No corresponding accumulation of weight is observed near $\mathcal{F}_1\approx 2$, despite two-site resonances being far more abundant at strong disorder~---~consistent with their vanishing contribution to $\mathcal{F}_1$. This provides a direct numerical confirmation that the minimal resonance generating non-Gaussianity involves $n=4$ different spins, consistent with the selection rule stated above.

\begin{figure}[t!]
    \centering
    \includegraphics[width=1\linewidth]{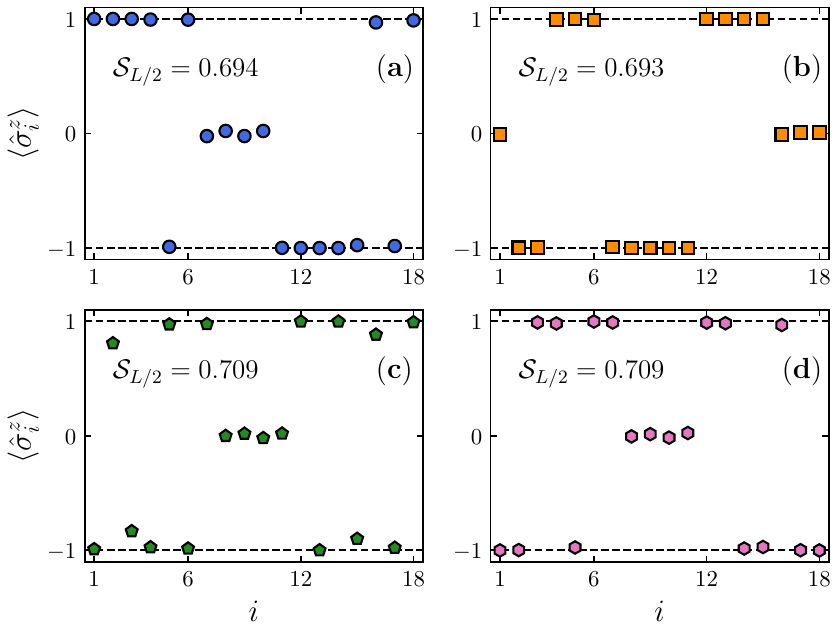}
    \caption{On-site spin magnetization profiles $\langle \hat{\sigma}^{z}_i\rangle$ of four different eigenstates with $\mathcal{F}_1 = 4\pm\epsilon$, where $\epsilon=10^{-3}$. The simulations are performed for $W=16$, $\Delta=1$, and $L=18$. The states follow the ansatz of Eq.~5 of the main letter, with four resonant spins and $L-4$ spins having frozen configurations. The half-chain entanglement entropy is close to $\mathcal{S}_{L/2}\approx\ln{2}$, providing additional evidence for the cat-like structure of such states.}    
    \label{Fig:Example_F4}
\end{figure}

The structure of these distributions closely parallels that of the entanglement entropy, where analogous peaks have been attributed to the existence of resonant eigenstates~\cite{Luitz16longTail}. Figure~\ref{Fig:Example_F4} illustrates the on-site magnetization profiles of four eigenstates obtained from different disorder realizations, selected such that $\mathcal{F}_1 =[4-\epsilon,4+\epsilon]$, with $\epsilon=10^{-3}$. In the numerical simulations, we consider $L=18$ and $W=16$, placing the system deep in the MBL regime. The selected states follow the structure of the ansatz described in Eq.~5: spins participating in the resonant state $\ket{\varphi}$ exhibit $\langle \hat{\sigma}_i^{z}\rangle\approx0$, while the remaining spins are practically frozen, with $\langle \hat{\sigma}_i^{z}\rangle\approx \pm{1}$. Moreover, the half-chain entanglement entropy is close to $\ln{2}$, providing strong evidence for the cat-like structure of these eigenstates. 

\subsection{Long-range cat-like eigenstates}

Although short-range resonances constitute the most common type of cat-like MBL eigenstates, rare long-range many-body resonances have attracted significant attention in recent years, as they have been argued to potentially destabilize the MBL regime~\cite{Villalonga20,Garratt21Local,Crowley21,colbois24interaction,Colbois24a,laflorencie25cat,morningstar22landmarks,Ha23}. In particular, system-wide resonances take the form of Eq.~5 of the main letter and, as shown in Ref.~\cite{laflorencie25cat}, can be identified through their spin-spin correlations. Specifically, their longitudinal connected correlator 
\begin{equation}
    C^{zz}_{i,i+L/2} = |\langle  \hat{\sigma}_{i}^{z}\hat{\sigma}_{i+L/2}^{z} \rangle - \langle \hat{\sigma}_{i}^{z}\rangle \langle \hat{\sigma}_{i+L/2}^{z}\rangle|  
\end{equation}
remains ${O}(1)$ for $W\gg1$, whereas typical MBL eigenstates have a vanishing connected correlator.

\begin{figure}[t!]
    \centering
    \includegraphics[width=1\linewidth]{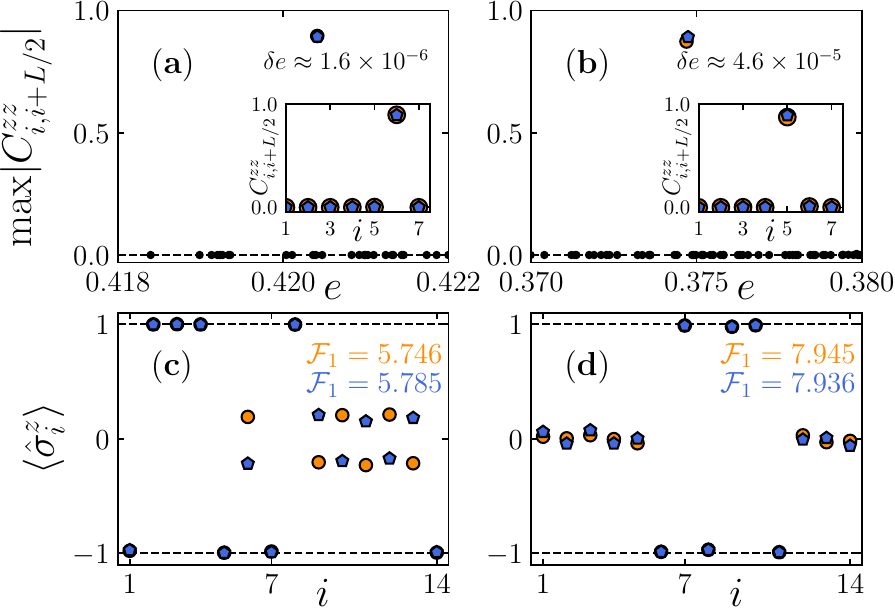}
    \caption{{Long-range cat eigenstates in the XXZ model for $W=16$ and $L=14$.} (a--b) The cat eigenstates exhibit significantly higher long-range connected longitudinal correlations $C_{i,i+L/2}^{zz}$ compared to typical MBL eigenstates and appear as quasi-degenerate pairs. (c--d) On-site magnetization $\langle \hat\sigma_{i}^{z}\rangle$ of the corresponding cat eigenstates shown in panels (a) and (b), respectively. The value of $\mathcal{F}_1$ provides a proxy for the number of spins participating in the cat-structure.}    
    \label{Fig:S4}
\end{figure}

Figure~\ref{Fig:S4} shows such rare cat eigenstates in the XXZ model for two different disorder realizations deep in the MBL regime ($W=16$), with $\Delta=1$ and $L=14$.  In panels (a-b), we show the maximal longitudinal connected correlators $C_{i,i+L/2}^{zz}$, maximized over all spin pairs $(i,i+L/2)$, for several eigenstates. Typical MBL eigenstates exhibit vanishing correlators, $C_{i,i+L/2}^{zz} \approx 0$, as indicated by the black dots. In contrast, a pair of eigenstates (colored markers) exhibits anomalously strong correlations, with $C_{i,i+L/2}^{zz} \approx 1$. These states appear as quasi-degenerate pairs, as evidenced by their small energy gap $\delta e$, where $e$ is the energy density. In the insets, we show the specific spin pair responsible for such high correlations, while the others remain close to zero.

In panels (c-d) of Fig.~\ref{Fig:S4}, we show the spin magnetization profile $\bra{\Psi_\mathrm{cat}^{\pm}}\hat\sigma_i^{z}\ket{\Psi_\mathrm{cat}^{\pm}}$ of these eigenstates. Panel~(c) corresponds to the cat eigenstates highlighted in panel~(a). Although the structure of $\ket{\Psi_\mathrm{cat}^{\pm}}$ does not correspond to an equal superposition of the hybridized product states, $\mathcal{F}_1$ remains bounded by the number of participating spins away from the exact resonance and indeed still provides a reliable proxy here, with $\mathcal{F}_1\approx 5.75-5.78$ for the two quasi-degenerate states shown. This relation is more visible in panel (d), which shows the magnetization profiles of the cat eigenstates displayed in panel (b): here the resonant spins follow an approximately equal superposition, with $\mathcal{F}_1\approx7.9$ for both states, closer to the true number of participating spins than in panel~(c).

As shown in Fig.~\ref{Fig:S4}, the number of spins participating in the long-range cat eigenstates differs from that of the short-range resonances. We now show that, unlike the short-range case, this number grows extensively with system size. In Fig.~17 of Ref.~\cite{laflorencie25cat}, the authors demonstrated that the number of fluctuating sites of $\ket{\Psi_\mathrm{cat}^{\pm}}$ increases approximately linearly with the system size, with a slope that depends on the disorder strength $W$. Since the fermionic antiflatness $\mathcal{F}_1$ provides a proxy for the number of fluctuating spins, its average behavior is expected to exhibit similar behavior.

\begin{figure}[t!]
    \centering
    \includegraphics[width=1\linewidth]{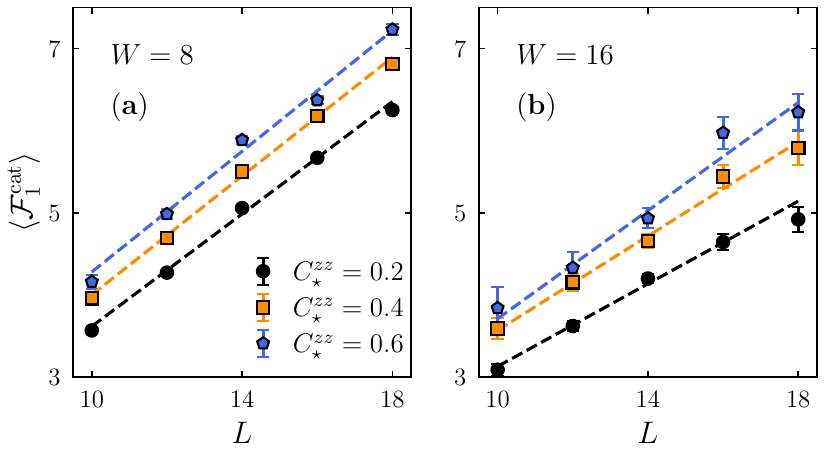}
    \caption{System-size dependence of the average value of the FAF $\langle \mathcal{F}_1\rangle$ of the long-range cat eigenstates for (a) $W=8$ and (b) $W=16$, and several values of the correlation threshold $C^{zz}_{\star}$. The average value grows approximately linearly with the system size, with the value of $\langle \mathcal{F}_1\rangle$ being comparable to the number of fluctuating sites in the cat-like eigenstates. }   
    \label{Fig:FAF_LR}
\end{figure}

In Fig.~\ref{Fig:FAF_LR}, we show the system-size dependence of the average FAF $\langle \mathcal{F}_1\rangle$ for the long-range cat eigenstates in the middle of the spectrum. These states are identified by selecting quasi-degenerate pairs of eigenstates with a longitudinal connected correlator $C^{zz}\geq C^{zz}_\star$, for three threshold values, $C^{zz}_\star = 0.2,0.4,0.6$. The average value of $\langle \mathcal{F}_1\rangle$ increases roughly linearly with $L$. For $W=8$, the slope of the linear scaling ($aL+b$) is approximately constant, with $a\approx0.36$. For $W=16$, the statistics are poorer due to the strong suppression of cat states, with the slopes depending more sensitively on the choice of $C^{zz}\geq C^{zz}_\star$. Nevertheless, the values of the fermionic antiflatness are comparable to the number of fluctuating sites reported in Ref.~\cite{laflorencie25cat}, further supporting the idea that the FAF is a proxy for the number of spins participating in the resonance.

\section{Resonant-pair channels and the mean FAF}~\label{Mechanism_Cats}

In the main letter, we numerically demonstrate that the \emph{average} value of $\mathcal{F}_1$ decays quadratically with disorder strength $W$, while the \emph{typical} value decays with $W^{-6}$, as shown in the End Matter and derived analytically in Sec.~\ref{Perturbative_Analysis}. This discrepancy arises because the mean is dominated by rare eigenstates, where the perturbative expansion fails. In this section, we show that a direct estimate of the leading rare event underestimates the observed scaling by one power of $W$, and that the missing power results from two adjacent resonant pairs connected by interaction on the bond between them.

We build on the two-level description presented in the previous section, which shows that the FAF of idealized cat-like eigenstates follows the Lorentzian behavior of Eq.~\eqref{Lorentzian_FAF}. Estimating the disorder-averaged value of these states requires integrating the Lorentzian over all values of the detuning $x=\delta E/2V$, rather than evaluating it at exact resonance. Averaging over different disorder realizations, we obtain the mean FAF of cat-like eigenstates,

\begin{equation}
    \mathbb{E}\left(\mathcal{F}_1^{\rm cat}\right) = n\rho(0) \cdot 2V\int_{-\infty}^{\infty}\frac{dx}{1+x^2} = 2\pi n\rho(0)V\,,
    \label{Average_FAF_cat}
\end{equation}
where $\rho(\delta E)$ is the detuning density. Since the resonance width $V$ is much smaller than the disorder scale $W$, the underlying uniform field distribution can be approximated as flat over the relevant window, allowing $\rho(0)$ to be treated as a constant. Analyzing the structure of the Hamiltonian in Eq.~\eqref{Interacting_Part}, we find that cat-like eigenstates formed via direct $\mathcal{U}^{(4)}$ channels yield $\mathcal{F}_1^{\rm cat} \propto \Delta/W^{3}$ (the $\mathcal{U}^{(3)}$ channel is excluded by the Gaussian-filter argument). While this explains the linear scaling with interaction strength $\Delta$, it does not account for the quadratic decay in $W$. We now demonstrate that this decay originates from pairs of resonant hybridizations rather than from a single direct process.

We now consider two resonant pairs located on bonds $(i-1,i)$ and $(i+1,i+2)$, where the fields are nearly aligned $|\delta h_1|,|\delta h_2|\leq J$, with $\delta h_1 = h_{i-1}-h_i$ and $\delta h_2 = h_{i+1}-h_{i+2}$. The central bond $(i,i+1)$ between them is off-resonant $|h_i-h_{i+1}|\gg J$. Since $\delta h_j$ is triangularly distributed on $[-2W,2W]$ with density approximately $1/2W$ near $\delta h_i\approx 0$, the probability of a resonant pair is $P(|\delta h_j|\leq J)\simeq J/W$ for $W\gg J$. 
As the two mismatches involve disjoint fields, the geometric probability is $P\left(|\delta h_1|,|\delta h_2|\leq J\right) \sim (J/W)^2$. The central-bond condition, $|h_i-h_{i+1}|\gg J$, holds with probability $1 - O(J/W)$, so it does not contribute at leading order.

The interaction on the central bond $(i,i+1)$ generates non-Gaussianity. In the strong disorder regime, the two resonant-pair subspaces are effectively decoupled from the remaining spins at leading order, allowing separate treatment to a first approximation. This description is further simplified by the block-diagonal structure imposed by U(1) symmetry. Among all sectors, only $N=2$ can yield nonzero FAF: the $N=0,4$ sectors correspond to product states, while the $N=1,3$ sectors are single-particle (one particle or one hole) and thus have Gaussian eigenstates.

The $N=2$~sector is six-dimensional, spanned by four pseudospin configurations $\left\{\ket{00}, \ket{01}, \ket{10}, \ket{11}\right\}$, where $\ket{0}=\ket{\bullet\circ}$ and $\ket{1} = \ket{\circ\bullet}$, as well as the states $\ket{\bullet\bullet\circ\circ}$ and $\ket{\circ\circ\bullet\bullet}$. The latter configurations are projected out via a second-order elimination, valid because they couple to the pseudospin manifold only via off-resonant single-particle hops with detuning of order $W$. This procedure shifts each pair's detuning by $O(J^2/W)$ but does not alter the pair-pair coupling. In the impurity model, the cluster remains isolated, rendering the description exact. In the XXZ Hamiltonian, the same pairs also border flanking interaction bonds, which modify $\delta h_i \rightarrow \delta h_i \pm \tfrac{\Delta}{2}\langle \hat{\sigma}^{z}_{\rm flank}\rangle$. This represents a second, independent correction that likewise shifts detunings without affecting the coupling. Therefore, to this order of approximation, the four-dimensional pseudospin manifold is governed by the effective Hamiltonian,

\begin{equation}
    \hat{\mathcal{H}}_{\rm eff} = \sum_{i=1,2} \tfrac{1}{2}\left(J\hat{\Gamma}_{i}^{x} + {\delta h_i}\hat{\Gamma}_{i}^{z}\right) -\lambda\hat{\Gamma}_{1}^{z}\hat{\Gamma}_{2}^{z}\,,
    \label{Effective_Hamiltonian_Subspace}
\end{equation}
where $\lambda=\Delta/4$ and $\hat\Gamma_i^z = +1\,(-1)$ indicate that the left (right) site of pair $i$ is occupied. All energy scales in $\hat{\mathcal{H}}_{\rm eff}$ are now $\lesssim J$; the interaction no longer competes against the bandwidth $W$, only against $J$.

In the non-interacting case ($\lambda=0$), each resonant pair is diagonalized as $\tfrac{1}{2}\left(J\hat{\Gamma}_{i}^{x} + {\delta h_i}\hat{\Gamma}_{i}^{z}\right) = \tilde{V}_{i}\tilde\Gamma_{i}^{z}$, with $\tilde{V}_i = (1/2)\sqrt{J^2 + \delta h_i^2}$ and eigenstates $\ket{\tilde{+}}_i = \cos\tfrac{\phi_i}{2}\ket{0} + \sin\tfrac{\phi_i}{2}\ket{1}$ and $\ket{\tilde{-}}_i = -\sin\tfrac{\phi_i}{2}\ket{0} + \cos\tfrac{\phi_i}{2}\ket{1}$, where $\tan\phi_i = J/\delta h_i$. The interaction couples the pairs through the operator rotation $\hat\Gamma^z_i = c_i\tilde\Gamma^z_i + s_i\tilde\Gamma^x_i$, with the full-angle amplitudes $s_i \equiv \sin\phi_i = J/\sqrt{J^2+\delta h_i^2}$ and $c_i \equiv \cos\phi_i = \delta h_i/\sqrt{J^2+\delta h_i^2}$. We now express the Hamiltonian~\eqref{Effective_Hamiltonian_Subspace} in the basis of the eigenstates of the non-interacting case ($\lambda=0$)
$\left\{\ket{\tilde{+}\tilde{+}},\ket{\tilde{+}\tilde{-}},
\ket{\tilde{-}\tilde{+}}, \ket{\tilde{-}\tilde{-}}\right\}$, where

\begin{equation}
    \hat{\mathcal{H}}_{\rm eff} = \begin{pmatrix}
        \Sigma\tilde{V}+ d_{\lambda} & b_2 & -b_1 & g \\
        b_2 & \delta\tilde{V} - d_{\lambda} & g & b_1 \\
        -b_1 & g & -\delta\tilde{V} - d_{\lambda} & -b_2 \\
        g & b_1 & -b_2 & -\Sigma\tilde{V} + d_{\lambda}
    \end{pmatrix}\,,
    \label{Effective_Hamiltonian_Subspace_v2}
\end{equation}
where $\Sigma\tilde{V} = \tilde{V}_1 + \tilde{V}_2$ and $\delta\tilde{V} = \tilde{V}_1 - \tilde{V}_2$. The remaining terms are $g = -\lambda s_1s_2$, $b_1 = \lambda s_1c_2$, $b_2 = \lambda c_1s_2$, $d_{\lambda} = -\lambda c_1c_2$.

The doublet states $\{\ket{\tilde+\tilde+},\ket{\tilde-\tilde-}\}$ have an energy gap of $2\Sigma\tilde{V}\sim2J$ and cannot form a resonance. In contrast, the doublet configuration $\{\ket{\tilde+\tilde-},\ket{\tilde-\tilde+}\}$ exhibits an energy splitting of $2\delta\tilde{V}$, which may vanish since $\tilde{V}_1$ and $\tilde{V}_2$ are independent. The single-flip elements $b_1$ and $b_2$ in Eq.~\eqref{Effective_Hamiltonian_Subspace_v2} couple the inner doublet to the outer doublet, split off by $2\Sigma\tilde V \sim 2J \gg \lambda$. Eliminating the outer doublet at second order, the two virtual paths (via $\ket{\tilde+\tilde+}$ and via $\ket{\tilde-\tilde-}$) cancel identically since these states sit symmetrically at $\pm \Sigma\tilde V$. The only effect is an unequal self-energy shift, $\Sigma_1 - \Sigma_2 = \lambda^2(s_1^2c_2^2/\tilde V_1 -c_1^2s_2^2/\tilde V_2)$, which vanishes at resonance. The coupling between these two states, $g=-\lambda s_1s_2 \sim O(\Delta)$, is generically nonzero, whereas the detuning density has $\rho(0)\sim1/J$ close to a perfect resonance. Restricted to the inner doublet, the effective Hamiltonian is then

\begin{equation}
    \hat{\mathcal{H}}_{\rm cats } = \begin{pmatrix}
        \delta \tilde{V} - d_{\lambda} & g \\
        g & -\delta \tilde{V} - d_{\lambda}
    \end{pmatrix}\,,
\end{equation}
which is precisely the two-level structure shown in the previous section, with detunings $\delta E = 2\delta \tilde{V}$ and coupling $V=|g|$. Away from resonance, the two-level result applies directly with $x=\delta E/2V$, yielding

\begin{equation}
    \mathcal{F}_1 = \frac{4}{1+x^2}\,,\qquad x = \frac{\delta \tilde{V}}{\lambda s_1 s_2}\,,
    \label{Cat_Collapse}
\end{equation}
with no additive contribution from $d_\lambda$, as it cancels identically from the detuning rather than only at resonance. At resonance ($\delta\tilde V=0$, $x=0$), the eigenstates are $\ket{\Psi^{\pm}_{\rm cat}} = \tfrac{1}{\sqrt{2}}\left(\ket{\tilde+\tilde-} \mp \ket{\tilde-\tilde+}\right)$, split by $2|g| = \Delta s_1s_2/2$, which remains independent of $W$, unlike the direct-channel $\mathcal{U}^{(4)}$. Expanding $\ket{\Psi^{\pm}_{\rm cat}}$ in the natural orbital basis, we have

\begin{equation}
    \ket{\Psi^{+}_{\rm cat}} = \sin\!\Big(\frac{\delta\phi}{2}\Big)\frac{\ket{00}+\ket{11}}{\sqrt2} + \cos\!\Big(\frac{\delta\phi}{2}\Big)\frac{\ket{01}-\ket{10}}{\sqrt2}\,,
    \label{Cat_Lab_plus}
\end{equation}
where $\delta\phi=\phi_1-\phi_2$. The resonance condition $\tilde V_1=\tilde V_2$ yields two branches: $\delta h_1=\delta h_2$ and $\delta h_1=-\delta h_2$. These branches are physically distinct in the laboratory basis, although both correspond to $\delta\tilde V=0$. In the first resonance condition, $\delta\phi=0$, the cat eigenstate $\ket{\Psi^{+}_{\rm cat}}$ is expressed as an equal superposition of two different lab basis states,

\begin{equation*}
    \ket{\Psi^{+}_{\rm cat}} \;\xrightarrow{\delta h_1=\delta h_2}\; \frac{\ket{\bullet\circ\circ\bullet}-\ket{\circ\bullet\bullet\circ}}{\sqrt2}\,.
\end{equation*}

In the second case, where $\delta h_1=-\delta h_2$, the cat eigenstate includes an admixture with the configurations $\{\ket{\bullet\circ\bullet\circ},\ket{\circ\bullet\circ\bullet}\}$, as $\sin(\delta\phi/2)=-c_1$. An analogous expansion applies to the second cat eigenstates. Defining $\Sigma\phi = \phi_1 + \phi_2$,

\begin{equation}
    \ket{\Psi^{-}_{\rm cat}} = -\sin\!\Big(\frac{\Sigma\phi}{2}\Big)\frac{\ket{00}-\ket{11}}{\sqrt2} + \cos\!\Big(\frac{\Sigma\phi}{2}\Big)\frac{\ket{01}+\ket{10}}{\sqrt2}\,.
    \label{Cat_Lab_minus}
\end{equation}

For this eigenstate, the resonance condition $\delta h_1 = -\delta h_2$ results in a pure two-configuration cat state,
\begin{equation*}
    \ket{\Psi^{-}_{\rm cat}} \;\xrightarrow{\delta h_1=-\delta h_2}\; \frac{\ket{\circ\bullet\circ\bullet}-\ket{\bullet\circ\bullet\circ}}{\sqrt2}\,,
\end{equation*}
whereas $\delta h_1 = \delta h_2$ produces an admixture with $\{\ket{\bullet\circ\circ\bullet}, \ket{\circ\bullet\bullet\circ}\}$. The two eigenstates are thus complementary: on either branch, exactly one of $\ket{\Psi^{\pm}_{\rm cat}}$ reduces to a clean two-configuration cat while the other is admixed, with the roles exchanged upon swapping $\delta h_1\leftrightarrow-\delta h_1$.

The Lorentzian collapse of Eq.~\eqref{Cat_Collapse} occurs regardless of which $\ket{\Psi^{\pm}_{\rm cat}}$ state is realized or the degree of admixture in its laboratory basis representation. The rotation connecting the laboratory basis to the non-interacting (tilde) eigenbasis is Gaussian, under which $\mathcal{F}_1$ remains invariant, regardless of the explicit admixture in the laboratory basis. The two eigenstates are nonetheless physically distinct and are resolved by the central-bond correlator $B_z\equiv\langle\hat\sigma^z_i\hat\sigma^z_{i+1}\rangle$. For the inner doublet, $B_z^{\rm inner}=c_1c_2\pm s_1s_2/\sqrt{1+x^2}$, with the sign corresponding to $\ket{\Psi^{\pm}_{\rm cat}}$, while the off-resonant (outer) doublet yields $B_z^{\rm outer}=-c_1c_2$. This correlator therefore provides a direct experimental means to determine which doublet the system occupies and, within the inner doublet, which of the two cat eigenstates has been selected.

The mechanism behind the quadratic decay with the disorder strength $W$ is now explained: the average FAF of the cats, given by Eq.~\eqref{Average_FAF_cat}, is independent of the disorder strength as $\rho(0)\sim1/J$ and $V=|g|$, with the $W^{-2}$ behavior arising from the probability of having two distinct resonances. Alternative mechanisms, such as a single resonant pair or a direct channel, exhibit faster decay rates in the idealized analysis, with $\mathcal{F}_1 \sim W^{-3}$. Consequently, the dimer channel provides the dominant contribution to the mean in the asymptotic regime.

We now contrast our theoretical predictions with the numerical results obtained from the microscopic Hamiltonian. For simplicity, we consider the impurity model $\hat{\mathcal{H}}_{\rm imp}$, where the interacting bond is fixed at $(i,i+1)=(L/2,L/2+1)$. This configuration places the two candidate resonant-pair bonds at $(L/2-1,L/2)$ and $(L/2+1,L/2+2)$. We filter the states by imposing that the occupancy of each resonant pair satisfies $|\langle \hat{\sigma}^{z}_{a}\rangle + \langle \hat{\sigma}^{z}_{a+1}\rangle| \leq \delta_{\rm occ}$ for $a\in\{L/2-1, L/2+1\}$, with $\delta_{\rm occ}=0.1$ as a threshold; similar results are observed for other choices of $\delta_{\rm occ}$. Additionally, we require the central and adjacent bonds to be off-resonant~---~$|h_{L/2} - h_{L/2+1}| \geq \delta_{\rm off-res}$, $|h_{L/2-2} - h_{L/2-1}| \geq \delta_{\rm off-res}$, and $|h_{L/2+2} - h_{L/2+3}| \geq \delta_{\rm off-res}$, with $\delta_{\rm off-res}$ several times $J$ (here, $\delta_{\rm off-res}=(5/2)J$)~---~so that the elimination of the outer levels applies both to the central bond and the pairs' external neighbors. The off-resonant condition does not affect the leading-order probability, so the fraction of realizations meeting all criteria is determined solely by the two resonant-pair probabilities and decreases as $(J/W)^2$. At accessible disorder strengths, these conditions introduce an additional, weakly $W$-dependent selection factor. Accordingly, at least $10^5$ disorder realizations are generated for each $(W,\Delta)$ to ensure sufficient statistical accuracy.

\begin{figure}[t!]
    \centering
    \includegraphics[width=1\linewidth]{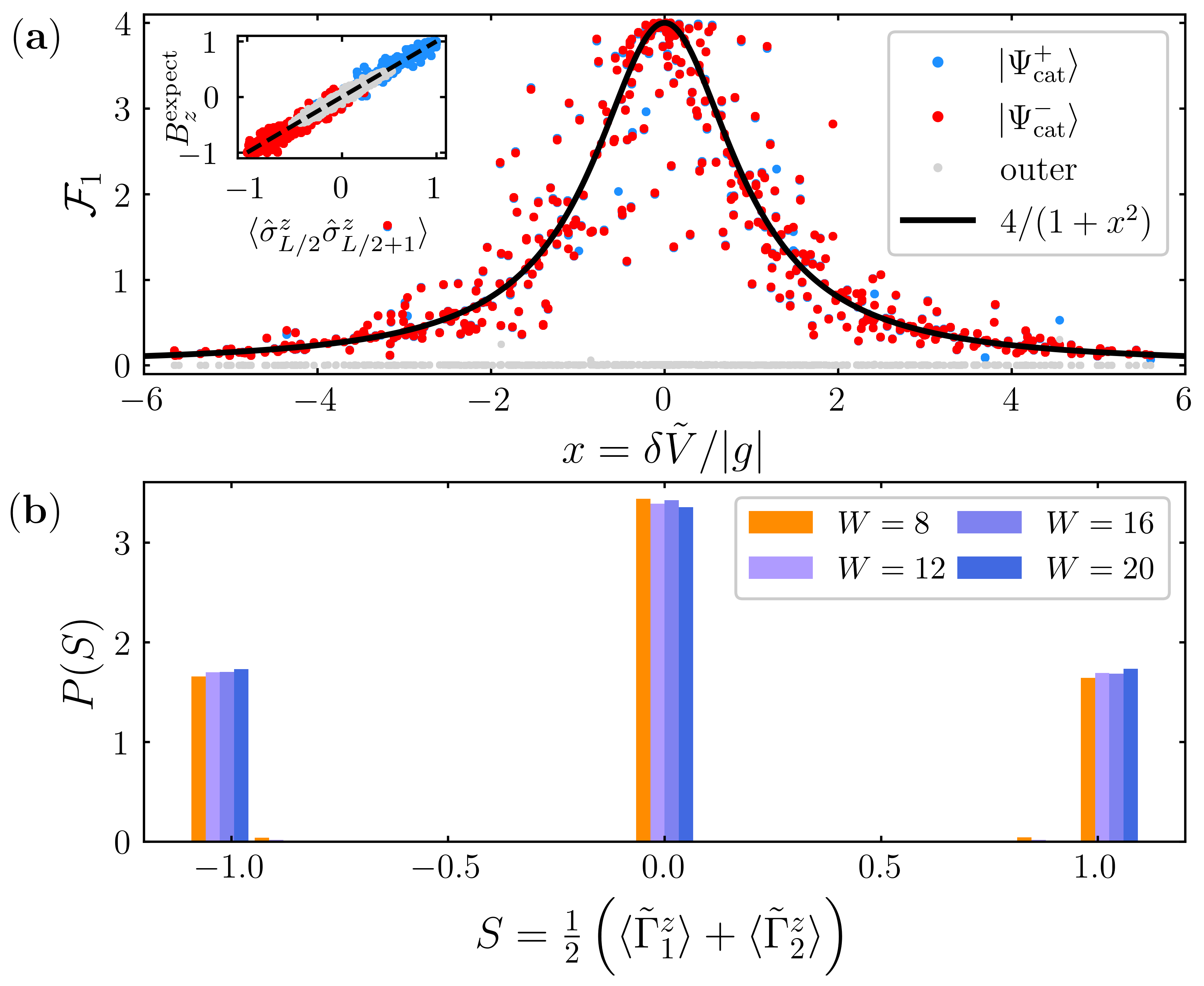}
    \caption{(a) FAF $\mathcal{F}_1$ as a function of $x=\delta \tilde V/|g|$ for all states surviving the resonant filter, at $\Delta=0.2$, $W=16$ and $L=14$. States in the inner doublet fall on the predicted Lorentzian (Eq.~\eqref{Cat_Collapse}; black curve), colored by which cat eigenstate they realize, $\ket{\Psi^{+}_{\rm cat}}$ (blue) or $\ket{\Psi^{-}_{\rm cat}}$ (red); states in the outer doublet (gray) show $\mathcal{F}_1\approx 0$. Inset: measured correlator $B_z\equiv\langle\hat\sigma^z_{L/2}\hat\sigma^z_{L/2+1}\rangle$ against the expected values $B_z^{\rm expect}$ for the inner doublet ($B_z^{\rm expect}=B_z^{\rm inner}$; see text) and outer doublet ($B_z^{\rm expect}=B_z^{\rm outer}$; see text).~(b) Trimodal distribution of $S = \tfrac{1}{2}\left(\langle \tilde\Gamma^{z}_{1}\rangle + \langle \tilde\Gamma^{z}_{2}\rangle  \right)$ for different disorder strengths $W$ at $\Delta=0.2$ and $L=14$. The eigenstates belonging to the \emph{inner} subspace have $S\approx0$ and are approximately $50\%$ of the states, whereas the other half belongs to the \emph{outer} subspace and has $S\approx \pm 1$. Bars for different $W$ are shown side by side within each peak for visual clarity.}
    \label{Fig_Cat_Resonant_1}
\end{figure}

In the main panel of Fig.~\ref{Fig_Cat_Resonant_1}(a), we show the FAF of all the states satisfying the filtering condition as a function of $x = \delta\tilde{V}/|g|$ for $\Delta=0.2$, $W=16$ and $L=14$. Two distinct behaviors emerge: for half of the states, $\mathcal{F}_1\approx0$ regardless of $x$~---~this class of states belongs to the outer doublet, whose energy splitting $2\Sigma\tilde V \sim 2J$ is far larger than its coupling $|b_i|\le \lambda/2$, well outside the resonance window described by $\hat{\mathcal{H}}_{\rm cats }$; on the other hand, the remaining states mimic the behavior expected for the Lorentzian (black solid curve), saturating near $\mathcal{F}_1\approx4$ at the resonance value $x\approx0$, and are further resolved into the two cat eigenstates $\ket{\Psi^{+}_{\rm cat}}$ (blue) and $\ket{\Psi^{-}_{\rm cat}}$ (red). In the inset, we plot the measured correlator $B_z\equiv\langle\hat\sigma^z_{L/2}\hat\sigma^z_{L/2+1}\rangle$ against the value predicted by $B_z^{\rm inner}$ for the inner-doublet states, and against $B_z^{\rm outer}$ for the outer-doublet states (gray). All three populations fall on the predicted diagonal, confirming that $B_z$ discriminates not only the inner/outer doublet split but, within the inner doublet, which of the two cat eigenstates has been selected.

A smoking gun signature of the mechanism is the trimodal distribution of the observable $S = \tfrac{1}{2}\left(\langle \tilde\Gamma^{z}_{1}\rangle + \langle \tilde\Gamma^{z}_{2}\rangle  \right)$. This observable differentiates which state the FAF probes: if $S \approx \pm 1$, the state belongs to the subspace of $\{\ket{\tilde{+}\tilde{+}}, \ket{\tilde{-}\tilde{-}} \}$, whereas if $S\approx 0$, the states belong to the inner subspace of $\{\ket{\tilde{+}\tilde{-}}, \ket{\tilde{-}\tilde{+}} \}$. The distribution shows three sharp peaks at the precise values, with the peak at $S\approx 0$ being twice as large as the other peaks. This behavior is obtained regardless of the disorder strength $W$ and interaction strength $\Delta$, confirming our analysis.

\begin{figure}[t]
    \centering
    \includegraphics[width=1.0\columnwidth]{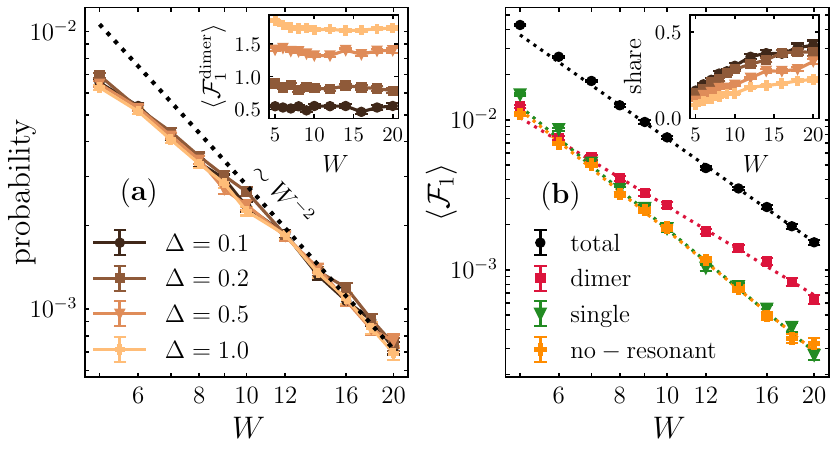}
    \caption{Statistics of the resonant pair states in the impurity model at $L=14$. (a) Probability that an eigenstate has two bonds of resonant fields and is singly occupied, for four values of $\Delta$; the collapse of the four curves shows the selection to be insensitive to the interaction. The black dashed line is a parameter-free prediction and decays as $W^{-2}$ at large disorder. Inset: mean FAF of the selected states, $\langle \mathcal{F}_1^{\rm dimer}\rangle$, as a function of $W$ for the same values of $\Delta$. It is flat in $W$~---~the entire disorder dependence of this channel lies in the geometry~---~and increases with the interaction strength $\Delta$. (b) Contribution to $\langle \mathcal{F}_1\rangle$ (black) from each channel at $\Delta=0.2$. Power laws are fitted for $W\geq8$. The dimer channel (red) decays with exponent $\gamma\approx 1.97(6)$, consistent with the predicted $2$, and grows in relative importance with $W$. The single-resonant-pair (green, $\gamma\approx2.72(6)$) and no-resonant-pair (orange,$\gamma\approx2.68(6)$) channels agree with each other, confirming that a single resonant pair confers no advantage over none. Inset: share of $\langle\mathcal{F}_1\rangle$ carried by the dimer channel, growing with $W$ for all $\Delta$ but not exceeding $1/2$ at accessible parameters.}
    \label{fig:cats_L14}
\end{figure}

We now test the probability of obtaining such states and their contribution to the mean FAF. Figure~\ref{fig:cats_L14}(a) presents the probability that an eigenstate has both bonds resonant and singly occupied, as a function of the disorder strength $W$ for four different values of $\Delta$ and $L=14$. The curves are $\Delta$-independent and approach a scaling $\propto W^{-2}$ at large disorder. The inset shows that the FAF conditioned on this manifold is flat in $W$ and increases with $\Delta$: the entire disorder dependence of this channel resides in the geometric probability, as the mechanism requires.

Figure~\ref{fig:cats_L14}(b) shows the contribution to the mean value of the FAF $\langle \mathcal{F}_1\rangle$ (black markers) from each mechanism. The contribution from dimers (red squares) decays more slowly at strong disorder than those from a single resonant pair (green) or from configurations with no resonance (orange), resulting in an increased relative importance with disorder. Fitting the different contributions with a power law, $\mathcal{F}_1 = CW^{-\gamma}$, for $W\geq 8$, we find $\gamma\approx 1.97(6)$ for the dimers, while $\gamma\approx2.72(6)$ and $\gamma\approx2.68(6)$ for the single and no-resonant configurations, respectively. The agreement between the latter two indicates that a single resonant pair does not provide an advantage over the absence of resonance. The difference in the extracted exponents demonstrates that, at strong disorder, the decay exponent of $\langle \mathcal{F}_1\rangle$ is determined by a specific geometric configuration. Applying the same fitting to $\langle \mathcal{F}_1\rangle$ yields $\gamma\approx2.29(3)$, which is consistent with a mixture of different channels. The inset of panel (b) indicates that the proportion contributed by the dimer channel increases with $W$ for all interaction strengths $\Delta$, but does not exceed $1/2$ within the studied range.

\section{Crossing point analysis}~\label{Crossing_Points}

\begin{figure}[t!]
    \centering
    \includegraphics[width=1.0\linewidth]{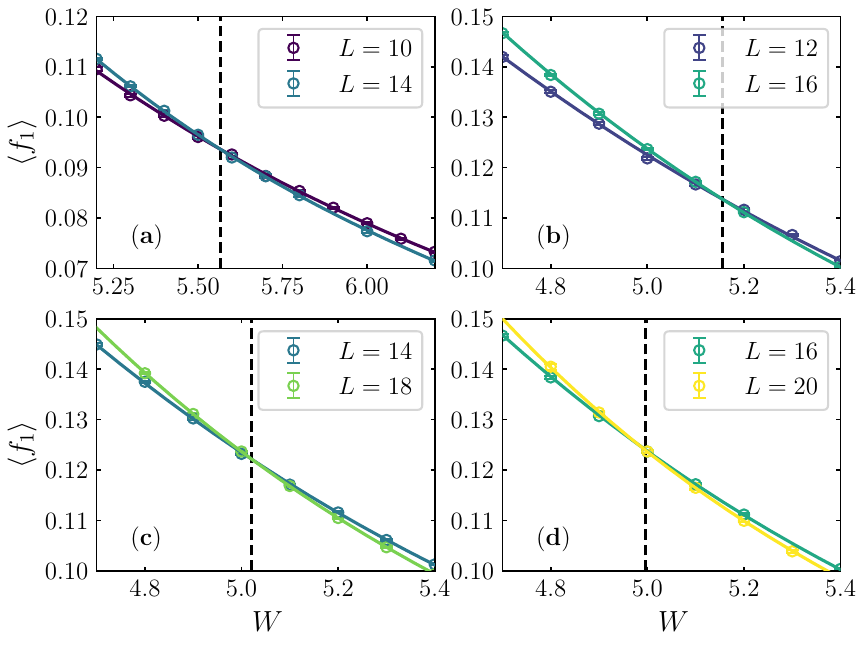}
    \caption{Crossing point analysis of the FAF in the disordered XXZ chain. Panels show $\langle f_1 \rangle$ as a function of disorder strength $W$ for (a) $L=10$ and $L=14$, (b) $L=12$ and $L=16$, (c) $L=14$ and $L=18$, (d) $L=16$ and $L=20$ . The crossing disorder strength $W_{\langle f_1\rangle}^{\star}$ \emph{decreases} with increasing system-size, appearing to saturate near $W\approx5$ in the largest system-size pairs analyzed. }
    \label{Fig:Crossing_Supmat}
\end{figure}

In the main letter, we show that the disorder-averaged FAF $\langle f_1\rangle$ for two different system sizes crosses at a disorder strength $W_{\langle f_1\rangle}^{\star}$ that is nearly independent of $L$, with $W_{\langle f_1\rangle}^{\star}\approx5$ for the XXZ Hamiltonian. In Fig.~5 of the main text, the flow of $W_{\langle f_1\rangle}^{\star}$ with system size is contrasted with that of the average gap ratio, a standard ergodicity-breaking indicator that exhibits pronounced finite-size drifts. Similar behavior is observed for other standard observables, such as the half-chain entanglement entropy or the Schmidt gap~\cite{Gray20Sgap}. In this section, we provide additional numerical details on how we extract the crossing points.

In Fig.~\ref{Fig:Crossing_Supmat}, we show the crossing of $\langle f_1\rangle$ between four pairs of different system sizes for the disordered XXZ Hamiltonian. To determine the crossing point accurately, we employ a large number of disorder realizations in the vicinity of the crossing region: $1.5\times10^{5}$ realizations for $L\leq18$ and at \emph{least} $3\times10^{4}$ realizations for $L=20$. We fit the disorder-averaged data with a third-order polynomial in $W$ over a small disorder window and identify $W_{\langle f_1\rangle}^{\star}$ as the intersection of the polynomial fits for each pair of system sizes; the extracted values are stable against variations of the fit window and of the polynomial order. In Fig.~\ref{Fig:Crossing_TypScaling}, $W_{\langle f_1\rangle}^{\star}$ is plotted against the mean size $\bar{L} = \tfrac{1}{2}(L_1 + L_2)$ of each pair.

Figure~\ref{Fig:Crossing_Supmat}(a) shows the crossing between $L=10$ and $L=14$, which occurs at $W\approx5.56$. This value is substantially larger than the crossing point of the gap ratio for the same pair of sizes (cf.\ Fig.~5 of the main letter). As the system size increases, the FAF crossing point flows toward \emph{smaller} disorder strength and appears to saturate near $W\approx 5$, in marked contrast to the behavior of other ergodicity-breaking indicators, whose crossing point continues to drift significantly across the accessible sizes.

\section{Additional analysis of the time-evolution}~\label{Sec.Time_Evolution}

In the main text, we show that expressing the FAF in terms of Majorana bilinear expectation values allows us to derive, within the $\ell$-bit framework (see End Matter), the dynamical behavior of $\mathcal{F}_1$,
\begin{equation}
    \mathcal{F}_1(\ket{\Psi_t}) = \mathcal{F}_1^{\rm MBL} - C\,t^{-\beta} + O(t^{-2\beta})\,,
    \label{Eq.Dynamical_Scaling}
\end{equation}
where $\mathcal{F}_1^{\rm MBL} \leq  B_{\infty}\equiv L - \sum_{k=1}^L{\braket{\hat{\sigma}_{k}^{z}}_{\infty}^2}$ is the saturation value and $\beta$ is an exponent that depends on the localization of the $\ell$-bits. Figure~\ref{Fig:Majoranas_FAF} shows qualitative agreement between the dynamics of Majorana expectation values in the XXZ model and those obtained in the $\ell$-bit framework. Panel~(c) further shows that the dynamics of $\braket{f_1}$ are consistent with Eq.~\eqref{Eq.Dynamical_Scaling}. Here, we examine the dynamical behavior of $\mathcal{F}_1$ under the XXZ dynamics across different system sizes, disorder strengths, and initial states, emphasizing the difficulty of distinguishing slow power-law relaxation from logarithmic growth over experimentally relevant time windows.

\subsection*{Validity of analytical scaling in the MBL regime}

\begin{figure}[t!]
    \centering
    \includegraphics[width=1\linewidth]{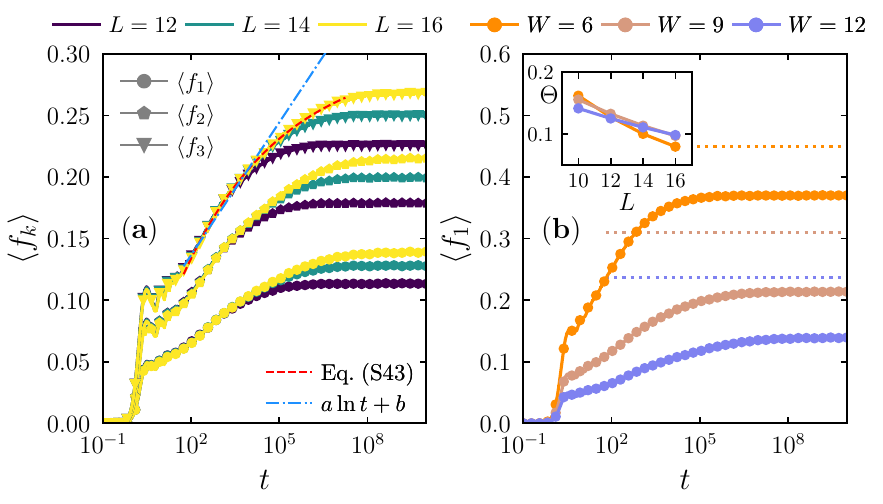}
    \caption{(a)~$\langle f_k\rangle$ dynamics as a function of $t$, shown on a logarithmic scale, for an initial N\'eel state $\ket{\Psi_{\mathrm{N\acute eel}}}= \ket{\uparrow\downarrow\ldots\uparrow\downarrow}$ for different system sizes and $k=1,2,3$. The red dashed line and the dash-dotted blue lines show fits to the power law of Eq.~\eqref{Eq.Dynamical_Scaling} and to a logarithmic growth $a\ln t+b$, respectively. (b) $\langle f_1\rangle$ dynamics for three disorder strengths, $W=6,9,12$, in the MBL regime, showing the consistency of our description; the colored dotted lines represent the upper bound $B_{\infty}/L$ for each disorder strength at $L=16$. Inset: system size scaling of the gap $\Theta$ (see text) for the three representative disorder strengths, showing that $\Theta$ decays monotonically with system size, suggesting the saturation value of $\braket{f_1}$ approaches the upper bound only at very large system sizes.}
    \label{Fig:FAF_scaling}
\end{figure}

Figure~\ref{Fig:FAF_scaling}(a) shows the dynamics of the disorder-averaged $\braket{f_k}\equiv \braket{\mathcal{F}_k}/L$ for $k=1,2,3$. We start from an initial N\'eel state $\ket{\Psi_{\mathrm{N\acute eel}}} = \ket{\uparrow\downarrow\ldots\uparrow\downarrow}$, evolving under the dynamics of the XXZ model in the MBL regime at $W=12$. We consider system sizes $L=12,14,16$ and use full exact diagonalization to access the evolution up to $10^{10}$ tunneling times. Following the initial transient time, the FAF exhibits slow growth and a gradual approach to saturation. Similar behavior is observed for all three indices $k$, although their saturation values differ.

Distinguishing the power-law relaxation described by Eq.~\eqref{Eq.Dynamical_Scaling} from the logarithmic growth constitutes a challenging task: for small $\beta$ and experimentally relevant times, one can expand $t^{-\beta}\approx t_0^{-\beta}\left[1-\beta\ln{(t/t_0)} + O(\beta^2\ln^2{(t/t_0)}) \right]$, where $t_0$ is a reference time. Therefore, when $|\beta\ln{(t/t_0)}|\ll1$, the leading relaxation term produces an approximately linear dependence on $\ln{t}$.

To illustrate this difficulty, Fig.~\ref{Fig:FAF_scaling}(a) compares fits based on Eq.~\eqref{Eq.Dynamical_Scaling} with logarithmic growth of the form $a\ln{t} + b$. The logarithmic fits reproduce part of the slow-growth regime but depart from the numerical data once the curvature toward saturation becomes visible. The power-law relaxation form of Eq.~\eqref{Eq.Dynamical_Scaling} remains consistent with the data over a broader time interval, with similar late-time departures from logarithmic growth observed across the system sizes and FAF indices considered. Thus, the numerical results favor the analytically predicted power-law approach to the saturation value, although the small exponent and limited accessible system sizes and time intervals do not permit a conclusive distinction from logarithmic growth. Distinguishing the two behaviors conclusively would require methods capable of treating larger system sizes while retaining access to asymptotically long times.

Figure~\ref{Fig:FAF_scaling}(b) shows the dynamics of $\braket{f_1}$ at $L=16$ for three representative values of disorder strength, $W=6,9,12$. The dynamics are qualitatively similar across these disorder strengths, with strong disorder suppressing the spread of fermionic magic resources and lowering the saturation value of $\langle f_1\rangle$ as $W$ increases. The colored dotted lines mark the disorder-averaged upper bound $B_{\infty}/L$; the visible gap between this bound and the saturated $\langle f_1\rangle$ shows that the $O(L^2)$ off-diagonal elements of the Majorana covariance matrix, though individually small, contribute collectively in a non-negligible way to the FAF. In the inset, we show the scaling with system size $L$ of the gap $\Theta\equiv B_{\infty}/L - \langle f_1^{\infty}\rangle$ between the upper bound and the saturation value of FAF for the same three disorder strengths $W$. We find that $\Theta$ decreases monotonically with $L$, showing that the saturation value of the FAF approaches the upper-bound prediction only at very large system sizes.

\subsection*{Robustness to the choice of initial state}

The numerical results in the main text and the previous subsection have so far been presented for the initial N\'eel state $\ket{\Psi_{\mathrm{N\acute eel}}}= \ket{\uparrow\downarrow\ldots\uparrow\downarrow}$. Here, we demonstrate the robustness of our analysis with respect to the choice of initial state, supporting its applicability to a broader class of initial Gaussian states in the MBL regime. In particular, we consider two classes of initial states: (a) random product states in the computational basis with energy near the middle of the many-body spectrum, denoted by $\ket{\Psi_{\rm rd}}$; and (b) an initial product state containing two domain walls, $\ket{\Psi_{\rm dw}} = \ket{\uparrow\cdots\uparrow\uparrow\downarrow\downarrow\cdots\downarrow}$~\footnote{The first domain wall is at the center of the chain; the second arises between the first and last sites due to periodic boundary conditions.}. 

The impact of the choice of initial state in the MBL regime was systematically investigated in~\cite{Haldar26}. This study identified a simple resonant mechanism that explains the controversies surrounding the imbalance dynamics~\cite{Sierant22c} and showed that the N\'eel state is rather special, exhibiting faster dynamics and a long-time imbalance that decreases with increasing system size. The domain-wall state $\ket{\Psi_{\rm dw}}$ is exceptional in the opposite sense, with the long-time imbalance increasing with $L$, consistent with localization.

\begin{figure}[t!]
    \centering
    \includegraphics[width=1\linewidth]{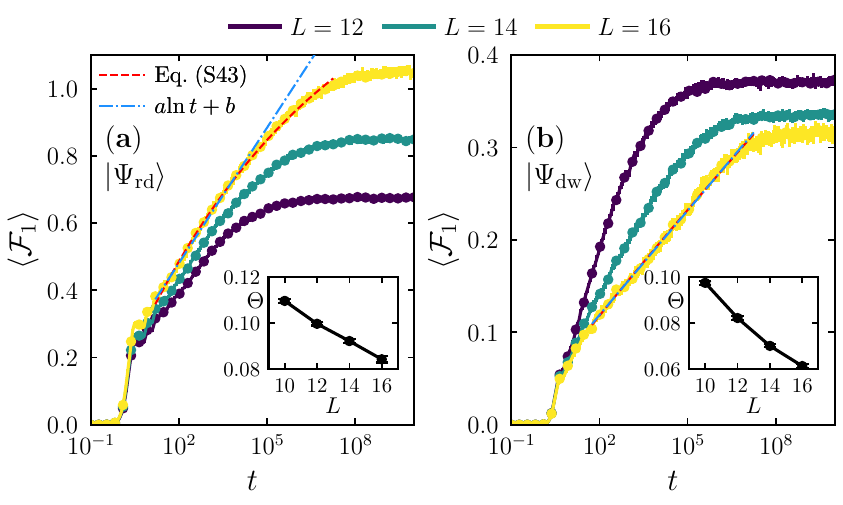}
    \caption{Time dynamics of $\langle \mathcal{F}_1\rangle$ deep in the MBL regime of the XXZ model for (a)~$\ket{\Psi_{\rm rd}}$ at $W=12$ and (b)~$\ket{\Psi_{\rm dw}}$ at $W=6$, at system sizes $L=12,14,16$. Red dashed and blue dash-dotted lines show fits to Eq.~\eqref{Eq.Dynamical_Scaling} and to $a\ln{t} + b$, respectively. Insets: system-size scaling of the gap $\Theta$ between the upper bound value $B_{\infty}/L$ and the true saturation value of $\langle f_1\rangle$, showing that $\Theta$ decreases monotonically with increasing $L$. }    
    \label{Fig:Additional_Results_TD}
\end{figure}

In Fig.~\ref{Fig:Additional_Results_TD}, we show the dynamics of $\langle \mathcal{F}_1\rangle$ deep in the MBL regime of the XXZ chain for (a) $\ket{\Psi_{\rm rd}}$ and (b) $\ket{\Psi_{\rm dw}}$, at several system sizes $L$. We set $\Delta=1$ and use periodic boundary conditions; however, our analysis is independent of the choice of boundary conditions. The numerical data are averaged over \emph{at least} $10^4$ disorder realizations. In panel~(a), we show the results for $W=12$. The growth of $\braket{\mathcal{F}_1}$ is well captured by Eq.~\eqref{Eq.Dynamical_Scaling} (red dashed line). Although the late-time curvature favors an approach to saturation, the small value of $\beta$ and the limited range of accessible system sizes and times prevent us from conclusively excluding logarithmic growth of the form $a\ln{t} + b$ (blue dashed-dotted line). 

In panel~(b), we consider a weaker disorder strength, $W=6$, since $\ket{\Psi_{\rm dw}}$ exhibits considerably slower dynamics than $\ket{\Psi_{\rm rd}}$ (or $\ket{\Psi_{\mathrm{N\acute eel}}}$) under the XXZ Hamiltonian, making it numerically challenging to resolve the long-time behavior at strong disorder. Here, $\braket{\mathcal{F}_1}$ \emph{decreases} with increasing $L$. This behavior reflects the structure of the initial state: the entangling dynamics is generated predominantly near the two domain walls, while the polarized domains remain essentially frozen over the accessible time scales. Consequently, the generated non-Gaussianity remains nonextensive. The residual decrease of $\braket{\mathcal{F}_1}$ with $L$ likely reflects finite-size effects associated with the increasing separation, and hence reduced dynamical overlap, between the two domain walls. In this case, it is particularly difficult to distinguish whether the prediction of Eq.~\eqref{Eq.Dynamical_Scaling} holds or logarithmic growth takes place. Because the power-law exponent $\beta$ is small, the two functional forms are effectively indistinguishable over the accessible system sizes and time intervals. The insets show the system-size dependence of the gap $\Theta$ for (a) $\ket{\Psi_{\rm rd}}$ and (b) $\ket{\Psi_{\rm dw}}$, confirming that $\Theta$ decreases monotonically with increasing $L$.

\section{Robustness of the results to the choice of FAF index}\label{Sec_FAF_index}

Throughout the main text and this supplementary material, we present results for $\mathcal{F}_1$. Here, we show that the choice of the index $k$ does not affect our conclusions. The FAF of order $k$ reads
\begin{equation*}
    \mathcal{F}_k = L - \tfrac{1}{2}\mathrm{tr}\left[\left(M^TM\right)^k\right]\,,
\end{equation*}
where $M$ is the $2L\times 2L$ covariance matrix and $L$ is the number of spins. The covariance matrix can be brought to its canonical block-diagonal form, and the FAF can be written in terms of the Williamson eigenvalues $\{\nu_{\alpha}\}$,

\begin{equation*}
    \mathcal{F}_k = \sum_{\alpha=1}^{L}\left(1 - x_{\alpha}^{k}\right)\,, \qquad x_{\alpha} = \nu_{\alpha}^2\,,
\end{equation*}
so that, for every $k$, $\mathcal{F}_k$ vanishes if and only if the state is Gaussian.

Each mode contribution of the FAF can be written as $\left(1 - x_{\alpha}^{k}\right) = (1-x_{\alpha})\sum_{m=0}^{k-1}x_{\alpha}^m$. Since $x_\alpha\in[0,1]$, we can write $\left(1 - x_{\alpha}\right)\leq\left(1 - x_{\alpha}^{k}\right)\leq k\left(1 - x_{\alpha}\right)$, arriving at the following inequality,

\begin{figure}[t!]
    \centering
    \includegraphics[width=1\linewidth]{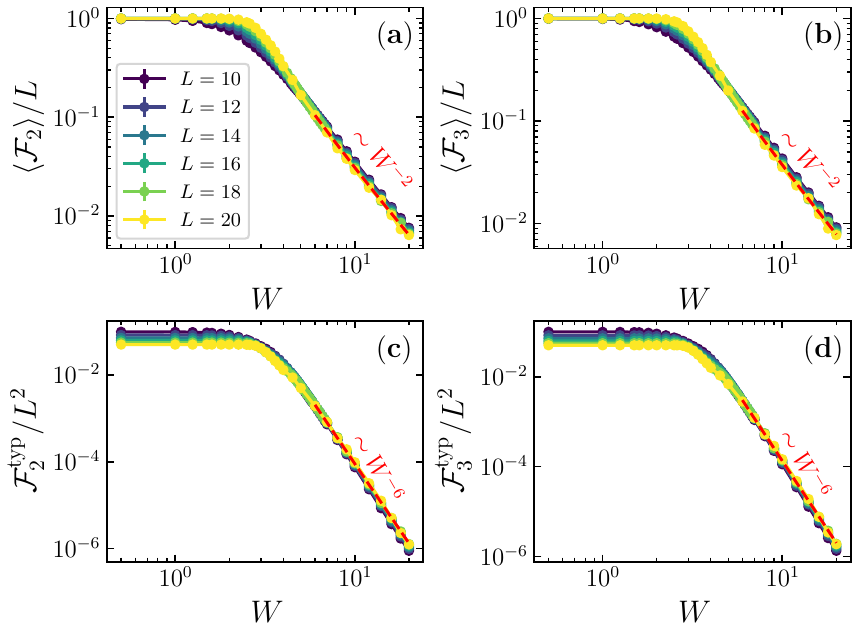}
    \caption{Upper panels: average FAF density $\braket{f_k}\equiv\braket{\mathcal{F}_k}/L$ as a function of the disorder strength $W$ for (a) $k=2$ and (b) $k=3$. Deep in the MBL regime, $\braket{f_k}\propto W^{-2}$ for any fixed $k\geq1$. Lower panels: rescaled median (typical) FAF, $\mathcal{F}_k^{\rm typ}/L^2$, for (c) $k=2$ and (d) $k=3$. The scaling $\mathcal{F}_k^{\rm typ}\propto L^2/W^6$ agrees with the main-text result for $\mathcal{F}_1^{\rm typ}$.}
    \label{Fig_S10}
\end{figure}

\begin{equation}\label{Inequality}
    \mathcal{F}_1\left(\ket{\Psi}\right) \leq \mathcal{F}_k\left(\ket{\Psi}\right) \leq k\mathcal{F}_{1}\left(\ket{\Psi}\right)\,,
\end{equation}
for any integer $k\geq1$. The lower bound is approached when $x_\alpha\to 0$, as for ergodic eigenstates, while close to Gaussianity ($x_\alpha = 1-\epsilon_\alpha$, $k\epsilon_\alpha\ll1$) one has $\mathcal{F}_k\simeq k\,\mathcal{F}_1$. Since Eq.~\eqref{Inequality} holds state by state, it carries over to disorder averages, $\braket{\mathcal{F}_1}\leq\braket{\mathcal{F}_k}\leq k\braket{\mathcal{F}_1}$, and to typical values, $\mathcal{F}_1^{\rm typ}\leq\mathcal{F}_k^{\rm typ}\leq k\,\mathcal{F}_1^{\rm typ}$. Consequently, for any fixed $k$, every asymptotic power-law dependence of $\mathcal{F}_1$ on $W$, $L$, or $\Delta$ is inherited by $\mathcal{F}_k$ with the same exponent, and only the prefactor can change, by at most a factor of $k$.

We confirm this numerically in Fig~\ref{Fig_S10}. The upper panels show the average FAF density, $\braket{f_k} \equiv \braket{F_k}/L$, as a function of the disorder strength $W$ for (a) $k=2$ and (b) $k=3$. In the ergodic regime, $\braket{f_k}\simeq1$ for all $k$. In the MBL regime, $\braket{f_k}$ decreases monotonically with $W$ and follows the scaling $\braket{f_k}\propto W^{-2}$ obtained for $k=1$ in the main text. The lower panels show the median (typical) FAF for (c) $k=2$ and (d) $k=3$, confirming that $\mathcal{F}_{k}^{\rm typ}\propto L^2/W^6$ deep in the MBL regime, as found for $\mathcal{F}_1^{\rm typ}$ in the main text.

The remaining results are likewise qualitatively robust with respect to the choice of $k$. For the cat-like eigenstates $\ket{\Psi^{\pm}_{\rm cat}}=\ket{\varphi}\otimes\ket{\Psi_{\rm fr}}$, the result follows solely from the additivity property of the FAF, which holds for any $k$, and from the fact that the covariance matrix of $\ket{\varphi}$ vanishes, making the contribution of $\ket{\varphi}$ independent of $k$. Regarding the dynamics, Fig.~\ref{Fig:FAF_scaling}(a) shows that, for an initial N\'eel state, $\braket{f_k(t)}$ with $k=2$ and $3$ displays the same qualitative behavior as $\braket{f_1(t)}$, as expected. This supports the conclusion that the main results presented throughout the manuscript remain valid for any fixed value of $k$.

\end{document}